\documentclass[]{article}
\usepackage[top=0.85in,left=2.75in,footskip=0.75in]{geometry}
\usepackage{fullpage}

\usepackage{amsmath,amssymb}

\usepackage{cite}

\usepackage{amsmath}
\usepackage{color}
\usepackage{amssymb}
\usepackage{graphicx}
\usepackage{fullpage}
\usepackage{epstopdf}
\usepackage{authblk}
\usepackage{nameref,hyperref}

\usepackage[table]{xcolor}

\usepackage{array}

\newcolumntype{+}{!{\vrule width 2pt}}

\definecolor{sion}{rgb}{.57,0.3,1}

\definecolor{lightblue}{rgb}{0,0,0}
\def\new#1{\textcolor{lightblue}{#1}}

\begin{document}

\title{Cliques and Cavities in the Human Connectome}

\author[1,2]{Ann Sizemore}
\author[1]{Chad Giusti}
\author[1]{Ari Kahn}
\author[1]{Richard F. Betzel}
\author[1,3,*]{Danielle S. Bassett}

\affil[1]{Department of Bioengineering, University of Pennsylvania, Philadelphia,
	PA 19041 USA}
\affil[2]{Broad Institute, Harvard University and the Massachusetts Institute of Technology, Cambridge, MA 02142 USA}
\affil[3]{Department of Electrical \& Systems Engineering, University of Pennsylvania, Philadelphia,
	PA 19041 USA} 
\affil[*]{To whom correspondence should be addressed: dsb@seas.upenn.edu}

\date{}
\maketitle
%
%




\section*{Abstract}
Encoding brain regions and their connections as a network of nodes and edges captures many of the possible paths along which information can be transmitted as humans process and perform complex behaviors. Because cognitive processes involve large and distributed networks of brain areas, principled examinations of multi-node routes within larger connection patterns can offer fundamental insights into the complexities of brain function. Here, we investigate both densely connected groups of nodes that could perform local computations as well as larger patterns of interactions that would allow for parallel processing. Finding such structures necessitates that we move from considering exclusively pairwise interactions to capturing higher order relations, \new{concepts} naturally expressed in the language of algebraic topology. \new{These tools can be used to study} mesoscale network structures that arise from the arrangement of densely connected substructures called \emph{cliques} in otherwise sparsely connected brain networks. We detect cliques (all-to-all connected sets of brain regions) in the average structural connectomes of 8 healthy adults scanned in triplicate and discover the presence of more large cliques than expected in null networks constructed via wiring minimization, providing architecture through which brain network can perform rapid, local processing. We then locate \emph{topological cavities} of different dimensions, around which information may flow in either diverging or converging patterns. These \new{cavities} exist consistently across subjects, differ from those observed in null model networks, and -- importantly -- link regions of early and late evolutionary origin in long loops, underscoring their unique role in controlling brain function. These results offer a first demonstration that techniques from algebraic topology offer a novel perspective on structural connectomics, highlighting loop-like paths as crucial features in the human brain's structural architecture.



\section*{Introduction}
Macroscopic computation and cognition in the human brain are affected by an intricately interconnected collection of neurophysical mechanisms \cite{bassett2010efficient,sporns2005human}. Unlike modern parallel computers, which operate through vast numbers of programs running in tandem and in isolation from one another, it is understood that many of these processes are supported on anatomically specialized brain regions that constantly share information among themselves through a network of white matter tracts \cite{hagmann2008mapping}. One approach for understanding the function of such a system begins with studying the organization of this white matter substrate using the language of networks \cite{sporns2015cerebral,bassett2011conserved,sporns2013human}. Collections of regions that are pairwise tightly interconnected by large tracts, variously known as communities \cite{porter2009communities}, modules \cite{meunier2009age}, and rich clubs \cite{van2011rich,senden2014rich}, have been the subject of substantial prior study. Moreover, they have given critical insights into the large-scale structural units of the brain that give rise to many common cognitive functions \cite{chen2008revealing,medaglia2015cognitive}. Such communities easily and rapidly transmit information among their members, facilitating local integration of information \cite{sporns2016modular}.

Often left implicit in such investigations of the white matter network is the understanding that just as important as the strong internal connections in communities are the relative \emph{weakness} of connections to external regions. This tendency to focus on strongly connected local regions arises naturally because standard network analyses are based on local properties of the network at individual vertices, where local edge strength is the primary feature \cite{bassett2006small,bullmore2009complex,bullmore2011brain}; the particular choice of quantitative language serves as a filter that diverts attention toward certain facets of the system. However, if one takes a more macro-scale view of the network, the small or absent white matter tracts intuitively serve to isolate processes carried on the strong white matter tracts from one another. Such structure facilitates more traditional conceptual models of parallel processing, wherein data is copied or divided into multiple pieces in order to rapidly perform distinct computations, and then recombined \cite{graham2011packet}. Together, the two notions of dense \new{cliques} and isolating cavities provide a picture of a system that performs complex computations by decomposing information into coherent pieces to be disseminated to local processing centers, and then aggregating the results.

In order to quantitatively characterize this macroscale structure, we employ an enrichment of networks that comes from the field of algebraic topology \cite{ghrist2014elementary}, developed precisely to understand the interplay between these weak and strong connections in systems \cite{ghrist2008barcodes}. Beginning with a structural white matter network, we first extract the collection of all-to-all connected subgraphs, called \emph{cliques}, which represent sets of brain regions that may possess a similar function, operate in unison, or share information rapidly \cite{sizemore2016classification}. Attaching these cliques to one another following a map given by the network creates a topological object called a \emph{clique complex} from which we can extract certain patterns of strongly connected regions called \emph{cycles} \cite{giusti2015clique}. \new{Chordless cycles} correspond to extended paths of potential information transmission along which computations can be performed serially to effect cognition in either a divergent or convergent manner\new{, and we refer to these ``enclosed spaces'' as \emph{topological cavities}\footnote{\new{In the mathematical literature, these are called \emph{non-trivial homology classes}. However, due to the extensive and potentially confusing collision with the use of the word ``homology'' in the study of brain function, here we elect to use this new terminology outside of references and necessary mathematical discussion in the Methods and Supplementary Information. Throughout, the word ``homology'' referes to the mathematical, rather than the biological, notion.}} in the network}. We hypothesize that the spatial distributions of cliques and \new{cavities} will differ in their anatomical locations, corresponding to their differential putative roles in neural computations. 

To address these hypotheses, we construct structural brain networks from diffusion spectrum imaging (DSI) data acquired from eight volunteers in triplicate. We measure node participation in cliques and compare these with a minimally wired null model \cite{betzel2016modular}. We also demonstrate the correspondence between the anatomical location of cliques and the anatomical location of the brain's structural rich club: a group of hubs that are densely connected to one another.  Next, we study \new{topological cavities} using a recently developed method from algebraic topology, which detects the presence and robustness \new{, summarized by a quantity called \emph{persistence),} of shell-like motifs of cliques called \emph{cycles}} in the network architecture. Specifically, we recover \new{all minimal length cycles corresponding to four persistent topological cavities} in the consensus structure, and show that these \new{features} are robustly present across subjects through multiple scans. Our results demonstrate that while cliques are observed in the structural core, \new{cycles enclosing topological cavities} are observed to link regions of subcortex, frontal cortex, and parietal cortex in long loops, underscoring their unique role in controlling brain function \cite{gu2015controllability,betzel2016optimally,muldoon2016stimulation}.



\section*{Results}

\begin{figure}[h]
	\centering
	\includegraphics[width = 3.5in]{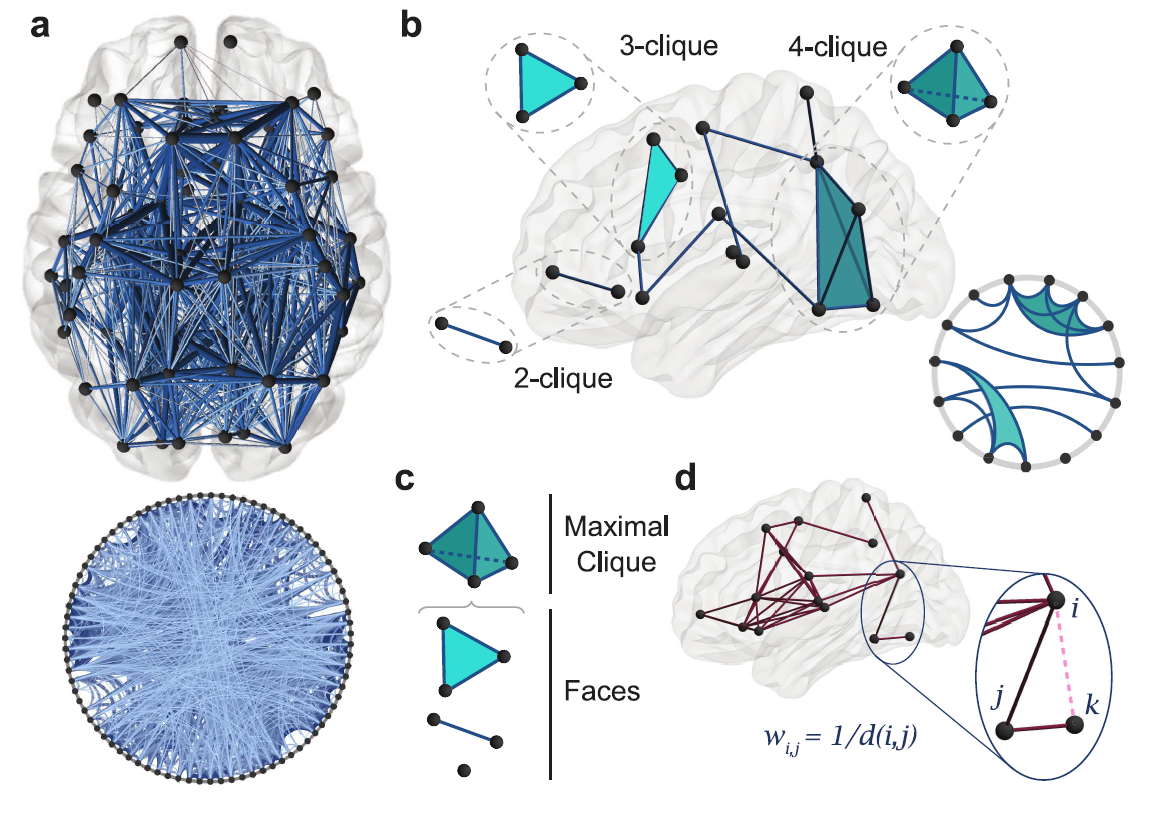}
	\caption{\textbf{Cliques are features of local neighborhoods in structural brain networks.} \emph{(a)} Diffusion spectrum imaging (DSI) data can be summarized as a network of nodes corresponding to brain regions, and weighted edges corresponding to the density of white matter streamlines reconstructed between them. Here we present a group-averaged network, where each edge corresponds to the mean density of white matter streamlines across eight subjects scanned in triplicate. \new{We show} the network at an edge density $\rho = 0.25$, and display its topology on the brain (\emph{top}), and on a circle plot (\emph{bottom}). \new{This and all brain networks are drawn with BrainNetViewer \cite{xia2013brainnet}.} \emph{(b)} All-to-all connected subgraphs on $k$ nodes are called $k$-cliques. For example, 2-, 3-, and 4-cliques are shown both as schematics and as features of a structural brain network. \emph{(c)} A maximal 4-clique has 3-, 2-, and 1-cliques as faces. \emph{(d)} For statistical validation, we construct a minimally wired null model by linking brain regions by edge weights equal to the inverse of the Euclidean distance between nodes \new{corresponding to brain region centers. Here we show an example of this scheme on 15 randomly chosen brain regions.}}
	\label{fig:1}
\end{figure}

To extract relevant architectural features of the human structural connectome, we first encoded diffusion spectrum imaging (DSI) data acquired from eight subjects in triplicate as undirected, weighted networks. In this network, nodes correspond to 83 brain regions defined by the Lausanne parcellation \cite{cammoun2012mapping} and edges correspond to the density of white matter tracts between node pairs (Fig.~\ref{fig:1}a). We initially study a group-averaged network, and then demonstrate that our results are consistently observed across individuals in the group as well as across multiple scans from the same individual.

\subsection*{\new{Cliques} in the Human Structural Connectome}~

\new{Here, we use the group-averaged network thresholded at an edge density ($\rho$) of 0.25 for computational purposes and for consistency with prior studies \cite{sizemore2016classification}. Results at other densities are similar, and details can be found in the Supplimentary Information. As a null-model, we use minimally wired networks (Fig.~\ref{fig:1}d) created from assigning edge weights to the inverse Euclidean distance between brain region centers (see Methods) observed in each of 24 scans. This model mimics the tendency of the brain to conserve wiring cost by giving edges connecting physically close nodes higher weight than edges between distant nodes.} 

\new{For each network, we now enumerate all maximal} $k$-cliques. \new{Recall that a $k$-clique is} a set of $k$ nodes having all pairwise connections (see Fig.~\ref{fig:1}b for 2-, 3-, and 4-cliques representing edges, triangles, and tetrahedra, respectively.) By definition, a subgraph of a clique will itself be a clique of lower dimension, called a \emph{face}. A \emph{maximal} clique is one that is not a face of any other (see Fig.~\ref{fig:1}c for a maximal 4-clique, which contains 3-, 2-, and 1-cliques as faces). 

\begin{figure*}[h!]
	\centering
	\includegraphics[width = \textwidth]{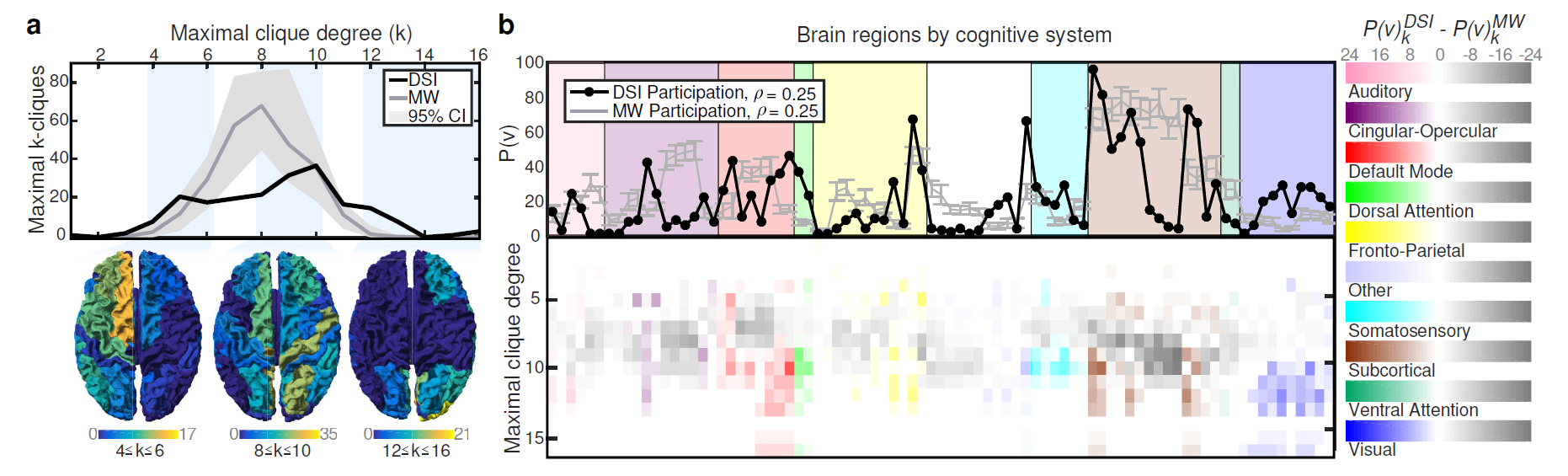}
	\caption{\textbf{Spatial distribution of maximal cliques varies between \new{average} DSI and minimally wired null model.} \emph{(a)} Distribution of maximal cliques in the \new{average} DSI (black) and \new{individual} minimally wired (gray) networks, thresholded at an edge density of $\rho=0.25$. Heat maps of node participation on the brain for a range of clique degrees equal to 4--6 (left), 8--10 (middle), and 12--16 (right). \emph{(b)} Node participation in maximal cliques sorted by the putative cognitive system to which the node is affiliated in functional imaging studies \cite{power2011functional}. We show individual node values (\emph{top}) as well as the difference between real and null model ($P_k^{DSI}-P_k^{MW}$; bottom) according to the colormap (\emph{right}). Individual node labels are listed in Fig. \ref{fig:sfig_table}.}
	\label{fig:fig2}
\end{figure*}

To understand the anatomical distribution of maximal cliques in both real and null model networks, we count the number of maximal $k$-cliques in which a node is a member, and refer to this value as the node participation, $P_k(v)$ (see Methods). Summing over all $k$ gives the total participation, $P(v)$. We observe that the distribution of maximal clique degrees is unimodal in the minimally wired null model and bimodal in the empirical data (see Fig.~\ref{fig:fig2}a). Anatomically, we observe a general progression of maximal clique participation from anterior to posterior regions of cortex as we detect higher degrees (Fig.~\ref{fig:sfig_corr}). Indeed, maximal cliques of 12--16 nodes contain nearly all of the visual cortex. This spatial distribution suggests that large interacting groups of brain regions are required for early information processing, while areas of frontal cortex driving higher-order cognition utilize smaller working clusters. We also observe that the human brain displays smaller maximal cliques than the minimally wired null model, a fact that might support its distributed processing.

The anterior-posterior gradient of maximal clique size can be complemented by additionally analyzing regional variation in the cognitive computations being performed. Specifically, we ask whether node participation in maximal cliques differs in specific cognitive systems \cite{power2011functional} (Fig.~\ref{fig:fig2}b).  We observe that the largest maximal cliques are formed by nodes located almost exclusively in the \new{subcortical, dorsal attention, visual, and default mode systems}, suggesting that these systems are tightly interconnected and might utilize robust topologically-local communication. Critically, this spatial distribution of the participation in maximal cliques differs significantly from the minimally wired null model, particularly in the cingulo-opercular and subcortical systems. We hypothesized that these differences may be driven by the excess of maximal 8-cliques in the minimally wired network (Fig.~\ref{fig:fig2}a). Expanding on the difference in node participation ($P_k^{DSI}(v) - P_k^{MW}(v)$), we see indeed that the large discrepancies between empirical and null model networks in cingulo-opercular and subcortical systems are caused by a difference in maximal cliques of approximately eight nodes (Fig.~\ref{fig:fig2}b, bottom).

\begin{figure}[h!]
	\centering
	\includegraphics[width = 3.5in]{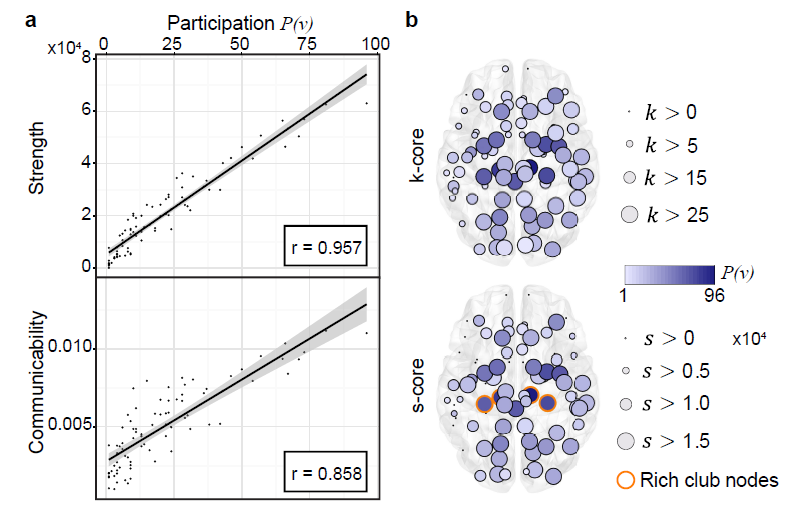}
	\caption{\textbf{Maximal clique participation tracks with network measures.} \emph{(a)} Scatter plot of node participation and node strength (\emph{top}) or communicability (\emph{bottom}). \emph{(b)} Calculated $k$-core (\emph{top}) and $s$-core decomposition in relation to participation in maximal cliques \new{with rich club nodes (shown $k_{RC}=43$; see Methods and Fig.~\ref{fig:sfig_richclub}) indicated in orange (\emph{bottom})}. Size indicates maximum $k$-core or $s$-core level attained by the node, while color indicates the participation $P(v)$.
	}
	\label{fig:fig3}
\end{figure}

A node with high participation must in turn be well connected locally\footnote{Note the converse is not necessarily true. As an example consider a node that only participates in one maximal 16-clique.}. Therefore we expect the participation of a node to act similarly to other measures of connectivity. To test this expectation, we examine the correlation of node participation with node strength, the summed edge weight of connections emanating from a node, as well as with node communicability, a measure of the strength of long distance walks emanating from a node (Fig.~\ref{fig:fig3}a). We find that both strength and communicability exhibit a strong linear correlation with the participation of a node in maximal cliques (Pearson correlation coefficient $r = 0.957$ and $r = 0.858$, respectively). 

These results indicate that regions that are strongly connected to the rest of the brain by both direct paths and indirect walks also participate in many maximal cliques. Such an observation suggests the possibility that brain hubs -- which are known to be strongly connected with one another in a so-called \emph{rich-club} -- play a key role in maximal cliques.  To test this possibility, we measure the association of brain regions to the rich-club using notions of coreness. A $k$-core of a graph $G$ is a maximal connected subgraph of $G$ in which all vertices have degree at least $k$, and an $s$-core is the equivalent notion for weighted graphs (see Methods). Using these notions, we consider how the $k$-core and $s$-core decompositions align with high participation (Fig.~\ref{fig:fig3}b). In both cases, nodes with higher participation often achieve higher levels in the $k$- and $s$-core decomposition. Moreover, we also observe the frequent existence of rich club connections between nodes with high participation (Fig.~\ref{fig:fig3}b, \emph{bottom}).  Together, these results suggest that rich-club regions of the human brain tend to participate in local computational units in the form of cliques.

\subsection*{Cavities in the Structural Connectome}~

\new{Whereas cliques in the DSI network act as neighborhood-scale building blocks for the computational structure of the brain, the relationships between these blocks can be investigated by studying the unexpected absence of strong connections, which can be detected as topological cavities in the structure of the brain network. Because connections are treated as communication channels along which brain regions can signal one another and participate in shared neural function, the absence of such connections implies a decreased capacity for communication which serves to enhance the segregation of different functions}. 

\begin{figure*}[]
	\centering
	\includegraphics[width = \textwidth]{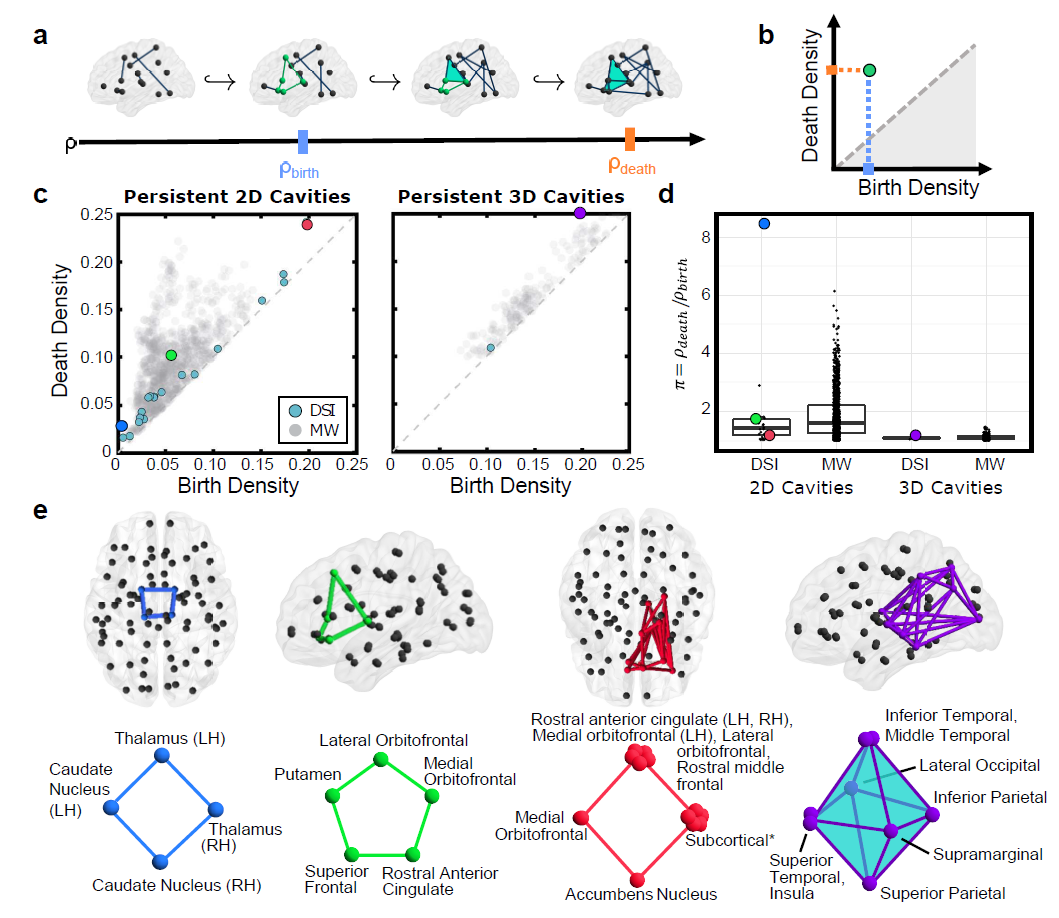}
	\caption{\textbf{Tracking clique patterns through a network filtration reveals \new{key topological cavities in the structural brain network}.} \emph{(a)} Example filtration of a network on 15 nodes shown in the brain across edge density ($\rho$). Blue line on the axis indicates the density of birth ($\rho_{birth}$) of the 2D cavity surrounded by the green minimal cycle. As edges are added, 3-cliques (cyan) form and shrink the cavity and consequentially the minimal green cycle is now four nodes in size. Finally, the orange line marks the time of death ($\rho_{death}$) when the cavity is now filled by 3-cliques. \emph{(b)} Persistence diagram for the cavity surrounded by the green cycle from panel \emph{(a)}. \emph{(c)} Persistence diagrams for the group-averaged DSI (teal) and minimally wired null (gray) networks in dimensions one (left) and two (right). Cavities in the group-averaged DSI network with long lifetime or high death-to-birth ratio are shown in unique colors and will be studied in more detail. \emph{(d)} Box plots of the death-to-birth ratio $\pi$ for cavities of two and three dimensions in the group-averagd DSI and minimally wired null networks. Colored dots correspond to those highlighted in panel \emph{(c)}. \new{The difference between $\pi$ values for \new{3D topological cavities} in the average DSI data versus the minimally wired null model was not found to be significant.} \emph{(e)} Minimal \new{cycles representing each persistent cavity at $\rho_{birth}$} noted in panels \emph{(c)}, \emph{(d)} shown in the brain (\emph{top}) and as a schematic (\emph{bottom}).}
	\label{fig:fig4}
	
\end{figure*}

\new{To identify topological cavities in a weighted network, we construct a sequence of binary graphs, each included in the next (Fig.~\ref{fig:fig4}a), known as a \emph{filtration}. Beginning with the empty graph we replace unweighted edges one at a time according to order of decreasing edge weight, and we index each graph by its \emph{edge density} $\rho$, given by the number of edges in the graph divided by the number of possible edges. After each edge addition, we \new{extract} ``shell-like'' motifs of $k$-cliques called (non-trivial) \emph{$(k-1)$-cycles}\footnote{\new{This shift in index is due to geometry: a $2$-clique is a 1-dimensional line segment, a $3$-clique is a 2-dimensional triangle, etc.}},  each of which encloses a $k$-dimensional topological cavity in the structure. When $k$ is clear or not pertinent, we will supress it from the notation, and refer simply to ``cycles'' and ``cavities''. While any cavity is surrounded by at least one cycle, often multiple cycles surround the same cavity. However, any two cycles that detect the same cavity will necessarily differ from one another by the boundaries of some collection of $(k+1)$-cliques (see Supporting Information and Fig.~\ref{fig:sfig_ph5}). Any two such cycles are called \emph{topologically equivalent}, so each topological cavity is detected by a non-trivial\footnote{\new{The equivalence class containing the cycle consisting of a single vertex is called \emph{trivial} and bounds the ``empty'' cavity.}}  \emph{equivalence class} of cycles.  We can represent a topological cavity using any of the cycles within the corresponding equivalence class, but for purposes of studying computational architectures it is reasonable to assume information will travel along paths of minimal length; thus, in this analysis we will consider the collection of cycles in an equivalence class with the minimal number of nodes\footnote{\new{In the absence of a filtration, there are serious computational issues involved in locating minimal-size representatives of equivalence classes. However, in this setting the computation is easily performed using standard algorithms (see Methods).}} and call these the \emph{minimal cycles} representing the cavity.} 

\new{As we move through the filtration by adding edges, the structure of the cycles, and thus of the cavities they represent, will evolve. We consider an example in Fig.~\ref{fig:fig4}a, showing a green minimal cycle surrounding a 2D cavity which first appears (is \emph{born}) in the graph sequence at $\rho_{birth}$ (cyan). As an edge completing a $3$-clique is added, the minimal cycle representative shrinks to four nodes in size, then finally is tessellated by $3$-cliques (\emph{dies}) at $\rho_{death}$ (orange). We record $\rho_{birth}$ and $\rho_{death}$ for all topological cavities (e.g., non-trivial equivalence classes of cycles) found within the filtration, and display them on a \emph{persistence diagram} (Fig.~\ref{fig:fig4}b). Cavities that survive many edge additions have a long \emph{lifetime}, defined as $\rho_{death} - \rho_{birth}$, or a large death-to-birth ratio, $\rho_{death}/\rho_{birth}$. Such cycles are commonly referred to as \emph{persistent cavities} and in many applications are considered the ``topological features'' of the system.}

\new{We investigate the persistence of 2D and 3D cavities (respectively represented by equivalence classes of $1$- and $2$-cycles) in the group-average DSI network and minimally wired null networks (see Fig.~\ref{fig:fig4}c). There are substantially fewer persistent cavities in the group-average DSI network than in the null models. To illustrate the structure of these cavities, we select four representative cavities with exceedingly long lifetimes or a high $\rho_{death}$ to $\rho_{birth}$ ratio (Fig.~\ref{fig:fig4}c,d) in the empirical data, and for each we find the minimal-length representative cycles at $\rho_{birth}$ (Fig.~\ref{fig:fig4}e)\footnote{\new{Such cycles for all of the persistent cavities found in the empircal data are illustrated in Fig.~\ref{fig:allcycles1} and \ref{fig:allcycles2}}}. The first persistent cavity appears as early as $\rho = 0.003$ and is minimally enclosed by the unique blue cycle composed of the thalamus and caudate nucleus of both hemispheres. The green cycle connecting the medial and lateral orbitofrontal, rostaral anterior cingulate, putamen, and superior frontal cortex is the only minimal cycle surrounding a long-lived cavity in the left hemisphere. The final persistent 2D cavity in the average DSI data is found in the right hemisphere between the medial orbitofrontal, accumbens nucleus, any of the subcortical regions hippocampus, caudate nucleus, putamen, thalamus, and amygdala, and any of the rostral middle frontal, lateral orbitofrontal, medial orbitofrontal of the left hemisphere, and rostral anterior cingulate from both hemispheres (see Fig.~\ref{fig:sfig_1cycles}e for all 12 minimal representatives). Finally, the purple octahedral cycle made from 3-cliques contains the inferior and middle temporal, lateral occipital, inferior parietal, supramarginal, superior parietal, and either of the superior temporal and insula of the left hemisphere, and encloses the longest-lived 3D cavity in the structural brain network.}

\subsection*{Test-ReTest Reliability and Other Methodological Considerations}~

\begin{figure*}[h!]
	\centering
	\includegraphics[width = \textwidth]{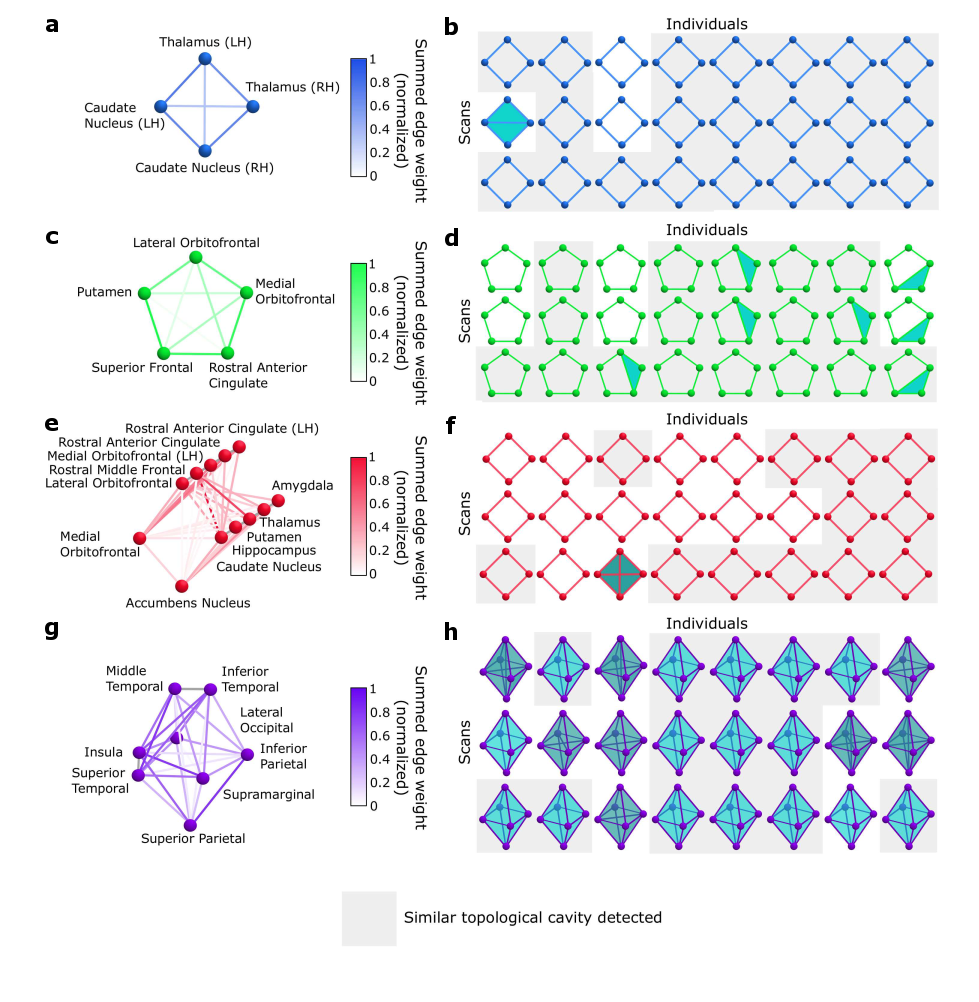}
	\caption{\textbf{Cycles \new{and similar cavities in the average DSI network} are consistently seen across individuals.} \new{\emph{(a),(c),(e),(g)} Normalized summed edge weights of the individual scan data connecting nodes seen in minimal cycle(s) recovered from average DSI. \emph{(b),(d),(f),(h)} Connections between indicated cycle nodes in each scan shown thresholded at the minimal weight of expected cycle edges. A gray background indicates a similar cavity found in this scan.}}
	\label{fig:sfig_1cycles}
	
\end{figure*}

It is important to ask whether the architectural features that we observe in the group-averaged DSI network can also be consistently observed across multiple individuals, and across multiple scans of the same individual \new{to ensure these cavities are not artifacts driven by a few outliers. Comparison of persistent cavities arising from two different networks is complicated by our notion of equivalence of cavities, and our desire to work with particular representative cycles. To capture the extent to which the cavities and their minimal representatives in the average DSI data are present in the individual scans, we record the collection of cliques that compose each minimal cycle representing the equivalance class (as seen in Fig.~\ref{fig:fig4}e), and check both for the existence of one of those collections of cliques, corresponding to the existence of the same strong fiber tracts, and, more stringently, for the presence of a topological cavity represented by that cycle in each individual's DSI network (see Supporting Information for more details). We observed that the subcortical cycle (Fig.~\ref{fig:fig4}e, blue) exists and these nodes (thalamus and caudate nucleus of both hemispheres) surround an equivalent 2D cavity in at least one scan of all individuals and the late-developing subcortical-frontal cycle (Fig.~\ref{fig:fig4}e, red) surrounds a cavity found in seven of the eight individuals in at least one of three scans (Fig.~\ref{fig:sfig_1cycles}b,f). The earlier arriving subcortical-frontal cycle (Fig.~\ref{fig:fig4}e, green) is present in all individuals and a similar cavity is seen at least once in all individuals (Fig.~\ref{fig:sfig_1cycles}d). Finally, we observe that the octahedral connection pattern in posterior parietal and occipital cortex (Fig.~\ref{fig:fig4}e, purple) is present at least once in seven of eight individuals and these regions enclose a similar cavity at least once in six of these individuals (Fig.~\ref{fig:sfig_1cycles}h). In the opposite hemisphere, the cyclic connection patterns and similar cavities appear though not as regularly (Fig.~\ref{fig:sfig_main4}). Finally we check the existence of similar cavities within the minimally wired null models, and see cavities denoted by the green and purple cycles are never seen (Fig.~\ref{fig:sfig_main4}). However, similar cavities to those represented by the red and blue minimal cycles appear frequently in the null model, though with different birth/death densities and lifetimes. In summary we find topological cavities observed in the group-averaged DSI network appear consistently across individuals, suggesting their potential role as conserved wiring motifs in the human brain.}

\begin{figure*}[h!]
	\centering
	\includegraphics[width = \textwidth]{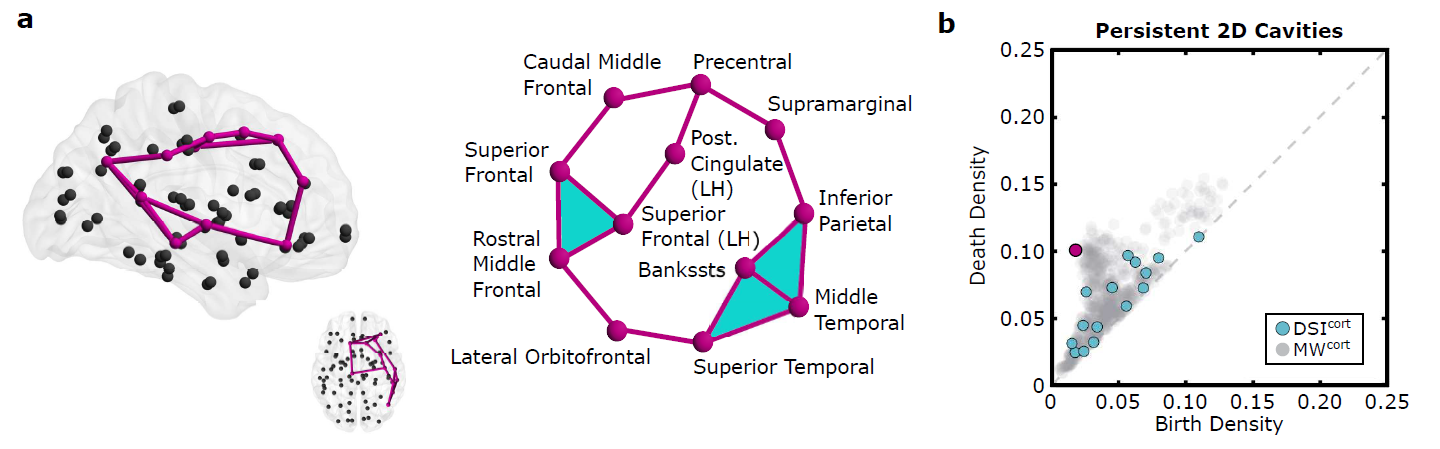}
	\caption{\textbf{Removal of subcortical \new{nodes allows for detection of nine-node cortical cycle enclosing large 2D cavity}.} \emph{(a)} \new{Minimal cycles} shown in the brain (\emph{left}) and as a schematic (\emph{right}). \emph{(b)} Persistence diagram of $DSI^{cort}$ and $MW^{cort}$. \new{Persistent feature corresponding to minimal cycles} in \emph{(a)} \new{indicated with maroon dot}.}
	\label{fig:fig6}
	
\end{figure*}

In addition to consistency across subjects and scans, it is important to determine whether the known high connectivity from subcortical nodes to the rest of the brain may be artificially \new{obscuring non-trivial} cortico-cortical \new{cavities} important for brain function. To address this question, we examined the 66-node group-average DSI network composed only of cortical regions, after removing subcortical regions, insula, and brainstem. We recovered \new{a long-lived topological cavity surrounded by four cycles of minimal length} composed of nine nodes connecting temporal, parietal, and frontal regions (Fig.~\ref{fig:fig6}). \new{Note in the schematic of Fig.~\ref{fig:fig6}a we see clearly two 2D cavities. The birth edge here was between the lateral orbitofrontal and superior temporal regions, which prevents us from determining whether the exact minimal cycle surrounding this cavity follows the superior frontal (LH)/posterior cingulate or the superior frontal (RH)/caudal middle frontal branch of the top loop. Following either of these two branches (then either of the banks of the superior temporal sulcus or middle temporal route) gives four cycles in which two are equivalent to each other but not to either cycle in the other pair. We will accept all of these four as minimal maroon cycles since any of the four could be minimal representatives}. Moreover, \new{at least one of these minimal cycles and corresponding cavity} was observed in each scan of every individual (Fig.~\ref{fig:sfig_1cyclesub}c), and often in the opposite hemisphere as well (Fig.~\ref{fig:sfig_1cyclesub}d). These results reveal that cortico-cortical cycles are indeed present and suggest their potential utility in segregating function across the brain.

\section*{Discussion}

In this study, we describe a principled examination of multi-node routes within larger connection patterns that are not accessible to network analysis methods that exclusively consider pairwise interactions between nodes. Our approach draws on concepts from a discipline of mathematics known as \emph{algebraic topology} to define sets of all-to-all connected nodes as structural units, called \emph{cliques}, and then to use the clique architecture of the network to detect structural topological cavities, \new{detected by the existence of non-trivial representative} \emph{cycles}. Using this approach, we show that node participation in maximal cliques varies spatially and by cognitive systems, suggesting a global organization of these neighborhood-scale features. These cliques form shell-like patterns of connectivity in the human structural connectome, which separate relatively early-evolving regions of the subcortex with higher-order association areas in frontal, parietal, and temporal cortex that evolved on more recent time scales. We found the recovered \new{topological cavities} exist consistently across individuals and are not expected in a spatially embedded null model, emphasizing their importance in neural wiring and function. These results offer a first demonstration that techniques from algebraic topology offer a novel perspective on structural connectomics, highlighting cavernous spaces as crucial features in the human brain's structural architecture.

\subsection*{Algebro-topological Tools for Neural Data Analysis}~

Algebraic topology is a relatively young field of pure mathematics that has only recently been applied to the study of real-world data. However, the power of these techniques to measure structures that are inaccessible to common graph metrics has gained immediate traction in the neuroscience community. Here, we highlight a few notable examples from the growing literature; a more comprehensive recent account can be found in \cite{giusti2016two}. At the neuron level, persistent persistence has been used to detect intrinsic structure in correlations between neural spike trains \cite{giusti2015clique}, expanding our understanding of the formation of spatial maps in the hippocampus \cite{dabaghian2012topological}. Moreover, at the level of large-scale brain regions, these tools have been exercised to characterize the global architecture of fMRI data \cite{stolz2014computational}.  Based on their unique sensitivity, we expect these algebric-topological methods to provide novel contributions to our understanding of the structure and function of neural circuitry across all scales at which combinatorial components act together for a common goal: from firing patterns coding for memory \cite{rajan2016recurrent,leen2015simple} to brain regions interacting to enable cognition.
 
\new{Our study uses algebraic topology in the classical form to obtain a global understanding of the structure, and in conjunction, it investigates particular topological features themselves and relates these features to cognitive function. Cycle representatives have previously been considered in biology \cite{chan2013topology,petri2014homological,lord2016insights}, but to our knowledge this is a first attempt to compare topological features in multiple brains.}

\subsection*{\new{Cliques} and Cavities for Computations}~

Cliques and \new{minimal cycles representing cavities} are structurally positioned to play distinct roles in neural computations. Cliques represent sets of brain regions that may possess a similar function, operate in unison, or share information rapidly \cite{sizemore2016classification}. Conversely, \new{minimal cycles} correspond to extended paths of potential information transmission along which computations can be performed serially to affect cognition in either a divergent or convergent manner.  Indeed, the \new{shell}-like or chain-like nature of cycles is a structural motif that has previously been -- at least qualitatively -- described in neuroanatomical studies of cellular circuitry. In this context, such motifs are known to play a key role in learning \cite{hermundstad2011learning}, memory \cite{rajan2016recurrent}, and behavioral control \cite{levy2001distributed,fiete2010spike}.  \new{The presence of minimal cycles suggests a possible role for polysynaptic connections and their importance to neural computations,} consistent with evidence from the field of computational neuroscience highlighting the role of highly structured circuits in sequence generation and memory \cite{rajan2016recurrent,hermundstad2011learning}. Indeed, in computational models at the neuron level, architectures reminiscent of chains \cite{levy2001distributed,fiete2010spike} and rings are particularly conducive to the generation of sequential behavioral responses. It is interesting to speculate that the presence of these \new{structures} at the much larger scale of white matter tracts could support diverse neural dynamics and a broader repertoire of cognitive computations than possible in simpler and more integrated network architectures \cite{tang2016structural}.

\new{Another consideration concerns the apparent asymmetry of our results with respect to left and right cerebral hemispheres. While unanticipated, we note that in some cases they have intuitive mathematical underpinnings. For example, in Fig.~\ref{fig:fig3}, we explicitly count maximal cliques, so one edge difference between a region in the left and right hemisphere could result in a large difference in the number of observed maximal cliques. Interestingly, despite this fact we still observe a strong correlation between node strength and $P(v)$, instilling confidence in these results. From a neuroscience point of view, brain asymmetries are not wholly unexpected. There is a storied and ever-growing literature describing the lateralization (i.e., asymmetries) of brain function \cite{galaburda1978right}. While speech generation \cite{rasmussen1977role} and language processing \cite{desmond1995functional,thulborn1999plasticity} are among the most commonly-cited functions to exhibit lateralization \cite{doron2012dynamic,chai2016functional}, such effects have also been linked to a diverse group of other cognitive domains. These include emotion \cite{wager2003valence}, processing of visual input \cite{sandi1993visual}, and even working memory \cite{carpenter2000working}. In addition, a number of studies have also reported the emergence of pathological lateralization or the disruption of asymmetries with neurocognitive disorders including ADHD \cite{oades1998frontal}. Our study does not offer a conclusive demonstration that the observed asymmetries arise from the lateralization of any specific brain function; we merely wish note that there is a precedent for such observations.}

\subsection*{Evolutionary and Developmental Drivers}~

Network filtration revealed several persistent \new{cavities} in the macroscale human connectome. While each \new{minimal} cycle \new{surrounding these cavities} involved brain regions interacting in a distinct configuration, we also observed commonalities across these structures. One such commonality was \new{these minimal} cycles tended to link evolutionarily old structures with more recently-developed neo-cortical regions \cite{rakic2009evolution}. For example, the green cycle depicted in Fig.~\ref{fig:fig4}e linked the putamen, an area involved in motor behavior \cite{middleton2000basal}, with \new{the rostral anterior} cingulate cortex, associated with higher-order cognitive functions such as error-monitoring \cite{braver2001anterior} and reward processing \cite{kringelbach2004functional}. This observation led us to speculate that the emergence of these \new{cavities} may reflect the disparate timescales over which brain regions and their circuitry have evolved \cite{gu2015emergence}, through the relative paucity of direct connections between regions that evolved to perform different functions. This hypothesis can be investigated in future work comparing the \new{clique and cavity structure} of the human connectome with that of non-human connectomes from organisms with less developed neocortices.

\subsection*{Towards a Global Understanding of Network Organization}
Though we highlighted \new{minimal} cycles in the brain, by nature persistence describes the global organization of the network. Often regions in the brain wire minimally to conserve wiring cost \cite{bassett2010efficient,bullmore2012,klimm2014resolving,lohse2014resolving}, though there are exceptions that give the brain its topological properties such as its small-world architecture \cite{bassett2006small,pessoa2014understanding,hilgetag2016brain,muldoon2016small,bassett2016small}. Following this idea, we could interpret the difference in \new{the number of persistent cavities} between the minimally wired and DSI networks as a consequence of the non-minimally wired edges, which \new{tessellate cavities} in the brain itself. Yet when the subcortical regions are removed, the persistent \new{cavities} of the minimally wired and DSI networks are much more similar (Fig.~\ref{fig:sfig_1cyclesub}b). This suggests that the wiring of cortical regions may be more heavily influenced by energy conservation than the wiring of subcortical regions. \new{Additionally the drop in the number and lifetime of persistent cavities when subcortical regions are included indicates that these subcortical regions may prematurely collapse topological cavities. The often high participation of subcortical regions in maximal cliques suggests these well-connected nodes may have hub-like projections to regions involved in cortical cycles, thus tessellating the cortical cavity with higher dimensional cliques.\footnote{Topologically these subcortical nodes are \emph{cone points}.} Previous studies have found that networks with ``star-like'' configurations are optimally efficient in terms of shortest-path efficiency, but also efficient in terms of a random walk-based measure of efficiency \cite{goni2013}. That is, networks optimized to have one or the other type of efficiency tend to have stars. Thus, stars appear to be useful configurations for fast communication, both along shortest paths and also in an unguided sense along random walks. The fact that we see star-like projections to cycles from subcortical regions may suggest that they are useful for efficient communication.}

\subsection*{Methodological Considerations}

An important consideration relates to the data from which we construct the human structural connectome. DSI and tractography, non-invasive tools for mapping the brain's white-matter connectivity, have some limitations. Tractography algorithms trade off specificity and sensitivity, making it challenging to simultaneously detect true connections while avoiding false connections \cite{thomas2014anatomical}, fail to detect superficial connections (i.e. those that do not pass through deep white matter)\cite{reveley2015superficial}, and have challenges tracking ``crossing fibers'', connections with different orientations that pass through the same voxel \cite{wedeen2008diffusion}. Nonetheless, DSI and tractography represent the only techniques for non-invasive imaging and reconstruction of the human connectome. While such shortcomings limit the applicability of DSI and tractography, they may prove addressable with the development of improved tractography algorithms and imaging techniques \cite{pestilli2014}.

\section*{Conclusion}

In conclusion, here we offer a unique perspective on the structural substrates of distinct types of neural computations. While traditional notions from graph theory and network science preferentially focus on local properties of the network at individual vertices or edges \cite{bassett2006small,bassett2009human,bullmore2009complex,bullmore2011brain}, here we utilize an enriched network formalism that comes from the field of algebraic topology \cite{ghrist2014elementary}. These tools are tuned to the interplay between weak and strong connections \cite{bassett2012altered}, and therefore reveal architectural features that serve to isolate information transmission processes \cite{giusti2016two}. It will be interesting in future to compare human and non-human connectomes across a range of spatial scales \cite{betzel2016multi} to further elucidate the evolutionary development of these features, and to link them to their functional \cite{hermundstad2013structural} and behavioral \cite{hermundstad2014structurally} consequences.

\section*{Materials and Methods}

\subsection*{Data Acquisition, Preprocessing, and Network Construction}
Diffusion spectrum imaging (DSI) data and T1-weighted anatomical scans were acquired from eight healthy adult volunteers on 3 separate days ($27\pm5$ years old, two female, and two left-handed) \cite{gu2015controllability}. All participants provided informed consent in writing according to the Institutional Review Board at the University of California, Santa Barbara. Whole-brain images were parcellated into 83 regions (network nodes) using the Lausanne atlas \cite{hagmann2008mapping}, and connections between regions (network edges) were weighted by the number of streamlines identified using a determistic fiber tracking algorithm. We represent this network as a graph $G(V,E)$ on $V$ nodes and $E$ edges, corresponding to a weighted symmetric adjacency matrix $\mathbf{A}$. For calculations in the main text, the network was thresholded at $\rho=0.25$ for consistency with previous work \cite{sizemore2016classification}. See Supporting Information and Refs \cite{cieslak2014local,gu2015controllability} for detailed descriptions of acquisition parameters, data preprocessing, and fiber tracking. In the supplement, we provide additional results for the case in which we correct edge weight definitions for the effect of region size \new{Fig.~\ref{fig:sfig_main4}}.

\subsection*{Cliques \emph{versus} Cycles}

In a graph $G(V,E)$ a $k$-clique is a set of $k$ all-to-all connected nodes. It follows that any subset of a $k$-clique is a clique of smaller degree, called a face. Any clique that is not a face we call maximal. To assess how individual nodes contribute to these structures, we define node participation in maximal $k$-cliques as $P_k(v)$, and we record the total participation of a node as $P(v) = \sum_{k = 1}^{n}P_k(v)$.

To detect cycles \new{which enclose topological cavities}, we computed the \emph{persistent homology} for dimensions 1--2 using \cite{henselmannovel}. This process involves first decomposing the weighted network into a sequence of binary graphs beginning with the empty graph and adding one edge at a time in order of decreasing edge weight. Formally, we translate edge weight information into a sequence of binary graphs called a filtration, $$ G_0 \subset G_1 \subset \dots \subset G_{|E|}$$ beginning with the empty graph $G_0$ and adding back one edge at a time following the decreasing edge weight ordering. \new{To ensure all edge weights are unique we added random noise uniformly sampled from $[0,0.0001]$. However, this has essentially no effect on the final results, as stability theorems ensure that small perturbation of the filtration leads to small perturbation of the persistent homology \cite{chowdhury2016persistent,cohen2007stability}.}

\new{Within each binary graph of this filtration, we extract the collection of all \emph{$k$-cycles}, families of $(k+1)$-cliques which, when considered as a geometric object, form a closed shell with no boundary. Formally, these are collections of $(k+1)$-cliques $\{\sigma_1, \dots \sigma_n\}$ such that every $k$-subclique of some $\sigma_i$ (called a \emph{boundary}) appears as a subclique in the collection an even number of times. Two $k$-cycles are \emph{equivalent} if they differ by a boundary of $k+1$-cliques.} This relation forms equivalence classes of cycles \new{with each non-trivial equivalence class representing a unique topological cavity}. 

Constructing the sequence of binary graphs allows us to follow \new{equivalence classes of cycles} as a function of the edge density $\rho$. Important points of interest along this sequence are the edge density associated with the first $G_i$ in which the \new{equivalence class} is found (called the birth \new{density}, $\rho_{birth}$) and the edge density associated with the first $G_i$ in which the enclosed void is triangulated into higher dimensional cliques (called the death \new{density}, $\rho_{death}$). \new{One potential marker of the} relative importance of a \new{persistent cavity} to the weighted network architecture is its \emph{lifetime} ($\rho_{death} - \rho_{birth}$). \new{A large lifetime indicates topological cavities} that persist over many edge additions. \new{An alternative measure is the} death to birth ratio $\pi = \rho_{death}/ \rho_{birth}$ which \new{highlights topological cavities that survive exceptionally long in spite of being born early, a feature that is interesting in geometric random graphs.} (see \cite{bobrowski2015maximally} and Supporting Information).

\new{To study the role of each topological cavity in cognitive function, we extract the minimal representatives of each non-trivial equivalence class at the birth density. For unfiltered complexes, the problem of finding a minimal generator for a given homology class is well known to be intractable \cite{chen2011hardness,dey2011optimal}. However, leveraging the filtration, we are able to answer the corresponding question in this context with relative ease. The persistent homology software \cite{henselmannovel} returns the starting edge and birth density of each homology class. To recover the minimal cycle we threshold the network at the density immediately preceding $\rho_{birth}$, then perform a breadth-first search for a path from one vertex to the other, taking all minimum length paths as solutions. If these minimum length paths arise from different equivalence classes (cycles 3, 7, and 10 in Fig.~\ref{fig:allcycles1}) we record and analyze each individually since each could be the generator of the homology class. For higher dimensional cycles we perform a similar process by hand, but we note that they could be algorithmically identified using appropriate generalizations of the graph search method.}

\subsection*{Standard Graph Statistics}
In addition to the notions of cliques and \new{cavities} from algebraic topology, we also examined corresponding notions from traditional graph theory including communicability and rich-club architecture, which are directly related to node participation in maximal cliques.

We first considered nodes that participated in many maximal cliques, and we assessed their putative role in brain communication using the notion of network communicability. \new{The weighted communicability between nodes $i$ and $j$ is $$C_{i,j} = (\exp(D^{-1/2}AD^{-1/2}))_{ij} $$ with $D := \text{diag}(s_i)$ for $s_i$ the strength of node $i$ in the adjacency matrix $\mathbf{A}$, providing a normalization step where each $a_{ij}$ is divided by $\sqrt{d_id_j}$ \cite{crofts2009weighted,estrada2008communicability}}. This statistic accounts for all walks between node pairs and scales the walk contribution according to the product of the component edge weights. The statistic also normalizes node strength to prevent high strength nodes from skewing the walk contributions. We refer to the sum of a node's communicability with all other nodes as node communicability, $C_{i}$.

Intuitively, nodes that participate in many maximal cliques may also play a critical role in the well-known rich club organization of the brain, in which highly connected nodes in the network are more connected to each other than expected in a random graph. For each degree $k$ we compute the weighted rich club coefficient $$\phi^w(k) = \frac{W_{>k}}{\sum_{l = 1}^{E_{>k}}w_l^{ranked}}$$ where $W_{>k}$ is the summed weight of edges in the subgraph composed of nodes with degree greater than $k$, $E_{>k}$ is the number of edges in this subgraph, and $w_l^{ranked}$ is the $l$-th greatest edge weight in $\mathbf{A}$. \new{Rich club nodes are those that exist in this subgraph when $\phi^w(k)$ is significantly greater (one sided $t$-test) than $\phi^w_{random}(k)$, the rich club coefficient calculated from 1000 networks constructed by randomly rewiring the graph $\mathbf{A}$ while preserving node strength \cite{rubinov2010complex}}.

Furthermore, highly participating nodes may also contribute to a hierarchical organization of the network. To evaluate this contribution, we compute the $k$-core and $s$-core decompositions of the graph \cite{hagmann2008mapping,chatterjee2007understanding}. The $k$-core is the maximally connected component of the subgraph with only nodes having degree greater than $k$. The $s$-core is similarly defined with summed edge weights in the subgraph required to be at least $s$.

\subsection*{Null Model Construction}
We sought to compare the empirically observed network architecture to that expected in an appropriate null model. Due to the well-known spatial constraints on structural brain \new{connectivity \cite{klimm2014resolving,lohse2014resolving,bullmore2012,betzel2016modular} as well as the similarity in mesoscale homological features to the Random Geometric network \cite{sizemore2016classification}} we considered a minimally wired network in which nodes are placed at the center of mass of anatomical brain regions. Each pair of nodes are then linked by an edge with weight $w_{i,j} = 1/d(i,j)$, where $d(i,j)$ is the Euclidean distance between nodes $i$ and $j$. \new{In each scan, the locations of region centers were collected. Thus, we considered a population of 24 model networks.} This null model allows us to assess what topological properties are driven by the precise spatial locations of brain regions combined with a stringent penalty on wiring length.

\section*{Acknowledgments}
This work was supported from the John D. and Catherine T. MacArthur Foundation, the Alfred P. Sloan Foundation, the Army Research Laboratory and the Army Research Office through contract numbers W911NF-10-2-0022 and W911NF-14-1-0679, the National Institute of Mental Health (2-R01-DC-009209-11), the National Institute of Child Health and Human Development (1R01HD086888-01), the Office of Naval Research, and the National Science Foundation (CRCNS BCS-1441502 and CAREER PHY-1554488). We thank Scott T. Grafton for access to the DSI data.

\bibliography{bibfile}
\bibliographystyle{plain}

\clearpage
\newpage

\section*{Supporting Information}


\section{Data Acquisition}
All participants volunteered with informed consent in writing in accordance with the Institutional Review Board/Human Subjects Committee of the University of California, Santa Barbara. Diffusion spectrum imaging (DSI) scans were acquired from eight subjects (mean age 27$\pm$5 years, two female, two left handed) on 3 separate days, for a total of 24 scans \cite{cieslak2014local}. DSI scans sampled 257 directions using a Q5 half-shell acquisition scheme with a maximum $b$-value of 5000 and an isotropic voxel size of 2.4 mm. We utilized an axial acquisition with the following parameters: repetition time (TR)=11.4 s, echo time (TE)=138 ms, 51 slices, field of view (FoV) (231,231,123 mm).

DSI data were reconstructed in DSI Studio (www.dsi-studio.labsolver.org) using $q$-space diffeomorphic reconstruction (QSDR) \cite{yeh2011ntu}. QSDR first reconstructs diffusion-weighted images in native space and computes the quantitative anisotropy (QA) in each voxel. These QA values are used to warp the brain to a template QA volume in Montreal Neurological Institute (MNI) space using the statistical parametric mapping (SPM) nonlinear registration algorithm. Once in MNI space, spin density functions were again reconstructed with a mean diffusion distance of 1.25 mm using three fiber orientations per voxel. Fiber tracking was performed in DSI studio with an angular cutoff of 55 degrees, step size of 1.0 mm, minimum length of 10 mm, spin density function smoothing of 0.0, maximum length of 400 mm and a QA threshold determined by DWI signal in the colony-stimulating factor. Deterministic fibre tracking using a modified FACT algorithm was performed until 100,000 streamlines were reconstructed for each individual.

In addition to diffusion scans, a three-dimensional high-resolution T1-weighted sagittal sequence image of the whole brain was obtained at each scanning session by a magnetization-prepared rapid acquisition gradient-echo sequence with the following parameters: TR=15.0 ms; TE=4.2 ms; flip angle=9 degrees, 3D acquisition, FOV=256 mm; slice thickness=0.89 mm, matrix=256 $\times$ 256. Anatomical scans were segmented using FreeSurfer \cite{dale1999cortical} and parcellated according to the Lausanne 2008 atlas included in the connectome mapping toolkit \cite{hagmann2008mapping}. A parcellation scheme including 83 regions was registered to the B0 volume from each subject’s DSI data. The B0 to MNI voxel mapping produced via QSDR was used to map region labels from native space to MNI coordinates. To extend region labels through the grey--white matter interface, the atlas was dilated by 4 mm. Dilation was accomplished by filling non-labelled voxels with the statistical mode of their neighbours’ labels. In the event of a tie, one of the modes was arbitrarily selected. Each streamline was labelled according to its terminal region pair.

\section{Additional neighborhood-scale computations}
\begin{figure}[h]
	\centering
	\includegraphics[width = \textwidth]{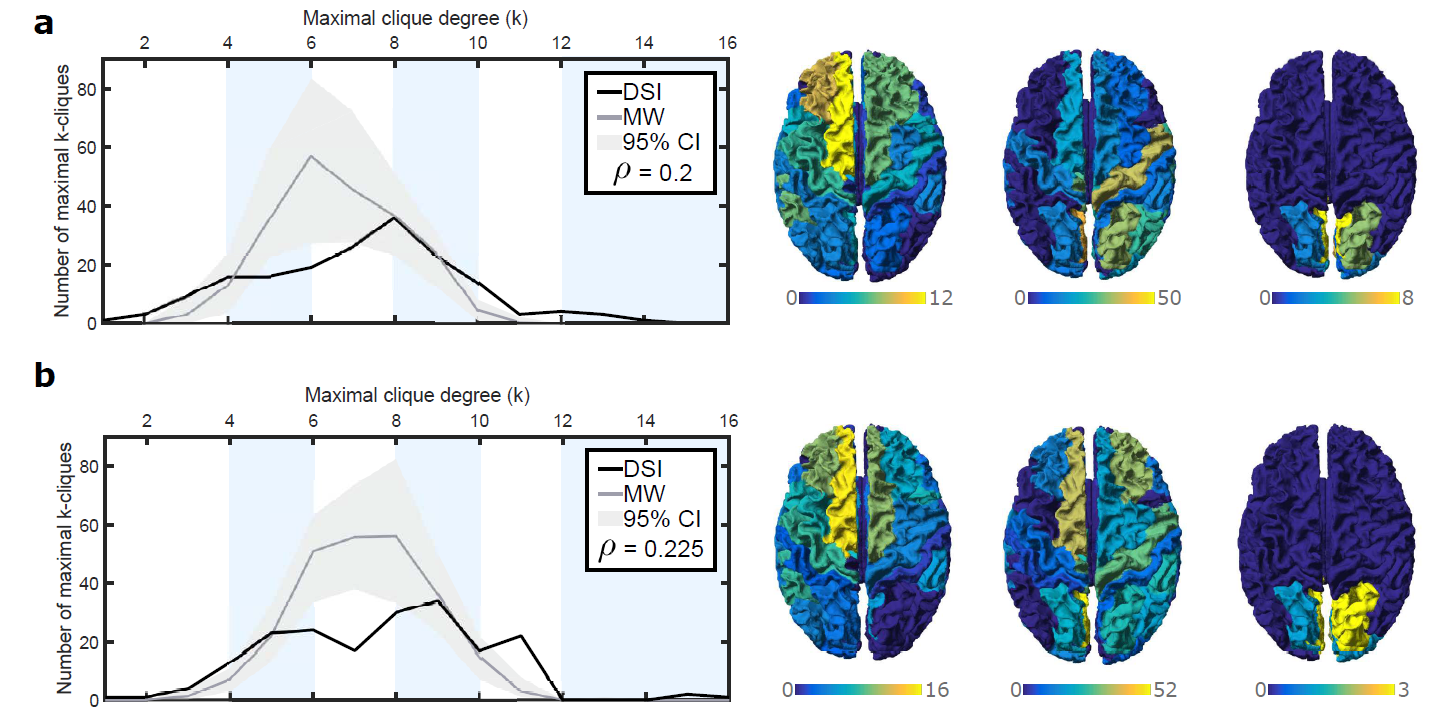}
	\caption{\new{Maximal clique distribution and spatial organization at multiple edge densities. \emph{(a)} Distribution of maximal cliques in DSI and minimally wired null model at $\rho = 0.2$ (left). Heat maps of node participation on the brain for maximal clique degrees 4-6 (left), 8-10 (middle), 12-16 (right). \emph{(b)} Plots as in \emph{(a)} at $\rho =0.225$.}}
	\label{fig:sfig_rhos}
\end{figure}

In the main text we count maximal cliques at an edge density of 0.25 (Fig.~\ref{fig:fig2}). To ensure our interpretation would not fluctuate based on this choice of $\rho$, we also show the maximal clique distribution for $\rho = 0.2$ (Fig.~\ref{fig:sfig_rhos}a) and $\rho= 0.225$ (Fig.~\ref{fig:sfig_rhos}b).

\begin{figure}[h!]
	\centering
	\includegraphics[width = 5in]{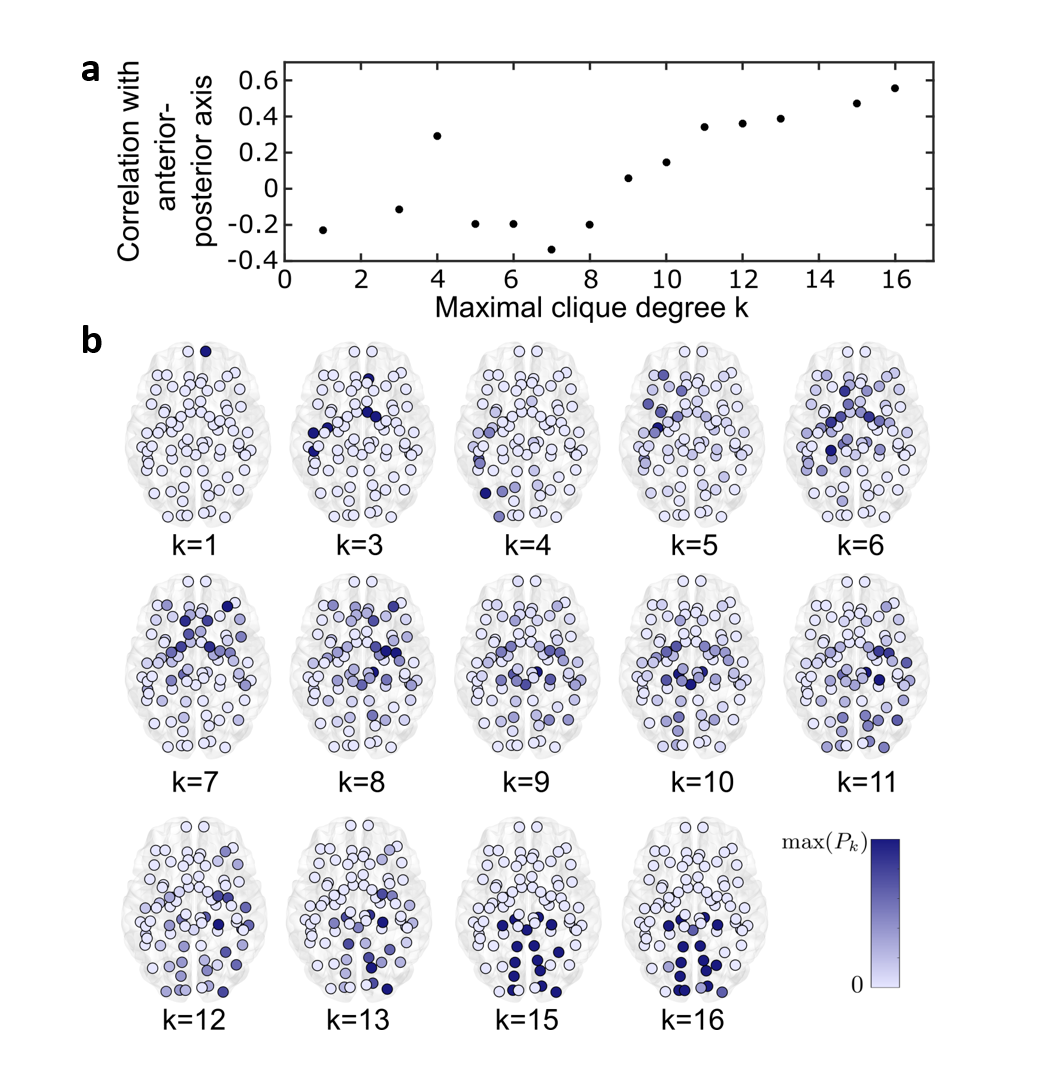}
	\caption{Maximal clique correlation with anterior-posterior gradient. \emph{(a)} Pearson correlation coefficient of $P_k(v)$ with the coordinate along the anterior-posterior axis. \emph{(b)} Spatial distribution of $P_k(v)$ for each $k$. Color of node corresponds to the value of $P_k(v)$ between zero and the maximum participation of any node for the given degree $k$.}
	\label{fig:sfig_corr}
	
\end{figure}

To address the extent to which an anterior-posterior gradient of maximal cliques exists, we calculated the correlation coefficient of $P_k(v)$ with the position of the node along this axis. Fig.~\ref{fig:sfig_corr} shows generally the maixmal participation of a node is more highly correlated with anterior-posterior position for higher degree cliques. To complement this calculation, Fig.~\ref{fig:sfig_corr}b shows the normalized $P_k(v)$ of each node for all maximal clique degree $k$. 

\begin{figure}[h!]
	\centering
	\includegraphics[height = 7in]{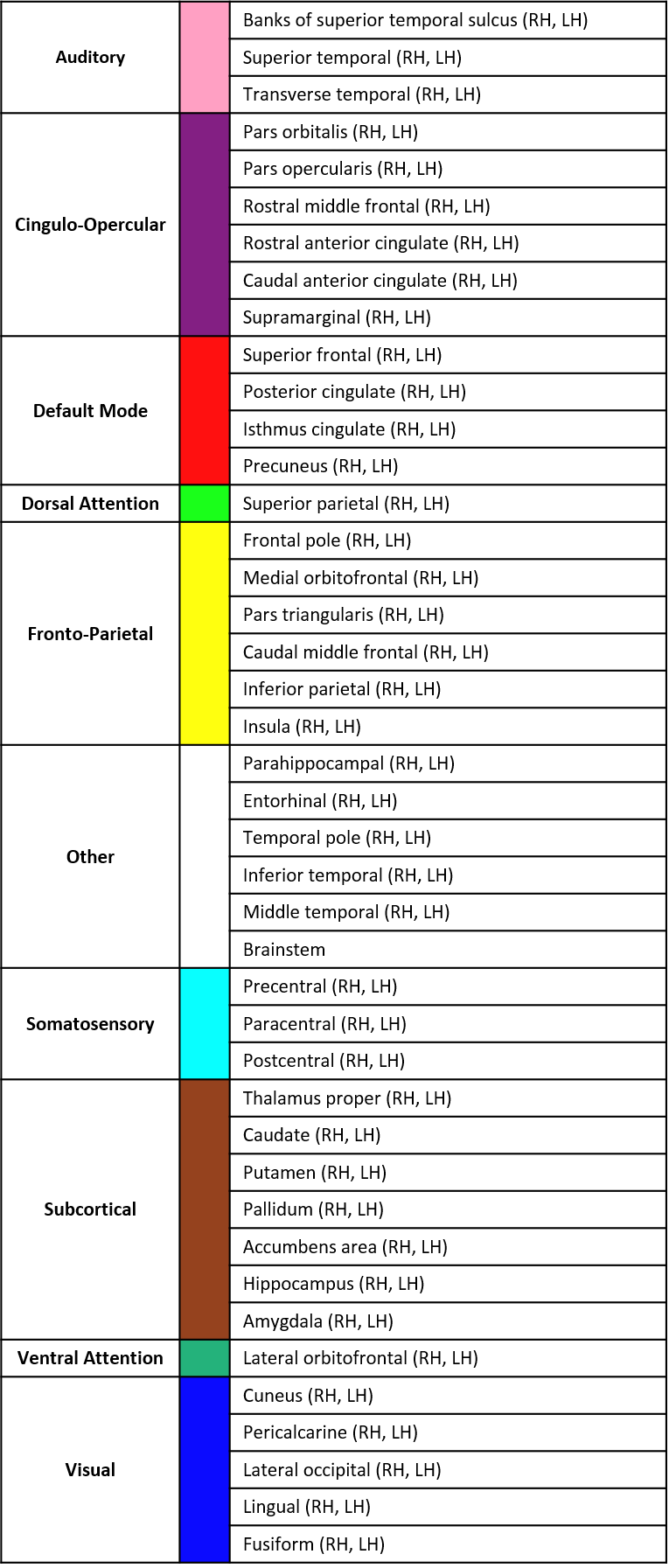}
	\caption{Order of brain regions for Fig.~\ref{fig:fig2}b. \label{fig:sfig_table}}	
\end{figure}

We then asked if node participation varies by cognitive system, perhaps reflecting each system's unique function. Results are shown in Fig.~\ref{fig:fig2}. The specific ordering of nodes for this figure are shown below (Fig.~\ref{fig:sfig_table}b). For each (right, left) hemisphere pair, the brain region in the right hemisphere was listed first, immediately followed by that in the left hemisphere.

\begin{figure}[h!]
	\centering
	\includegraphics[width = 4in]{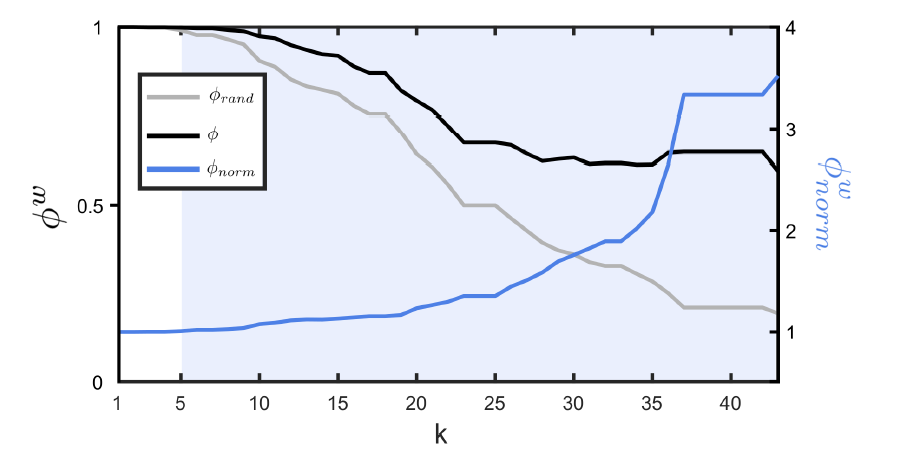}
	\caption{Defining the rich club of the DSI network. Rich club coefficient of the DSI network ($\phi(k)$) is shown in black, the average rich club coefficient of randomized networks ($\phi_{rand}(k)$) in gray, and the normalized rich club coefficient $\phi_{norm}(k)$) in blue. Shaded regions indicate values of $k$ for which $\phi(k)$ significantly exceeded $\phi_{rand}(k)$.
		\label{fig:sfig_richclub}}

\end{figure}

Additionally we are interested in comparing node participation to other measures of connectedness, as we expect they should generally agree. One such measure is the rich club. Following the procedure of van den Heuvel and Sporns \new{\cite{van2011rich}}, we calculated $\phi$, $\phi_{rand}$, and $\phi_{norm}$ for each value of $k$ (Fig.~\ref{fig:sfig_richclub}).

\clearpage
\newpage

\section{Persistent Homology}

We are interested in finding mesoscale structural features, specifically \new{\emph{non-trivial minimal cycles}} within our weighted network. Persistent homology strings together these features across network snapshots in a \emph{filtration}, offering a global \new{picture} of network architecture. We include a brief description of the method here, and we advise the interested reader to consult \cite{carlsson2009topology,ghrist2014elementary,zomorodian2005computing} for additional details.

\subsection{Complexes}
\begin{figure}[h]
	\centering
	\includegraphics[width = 3.5in]{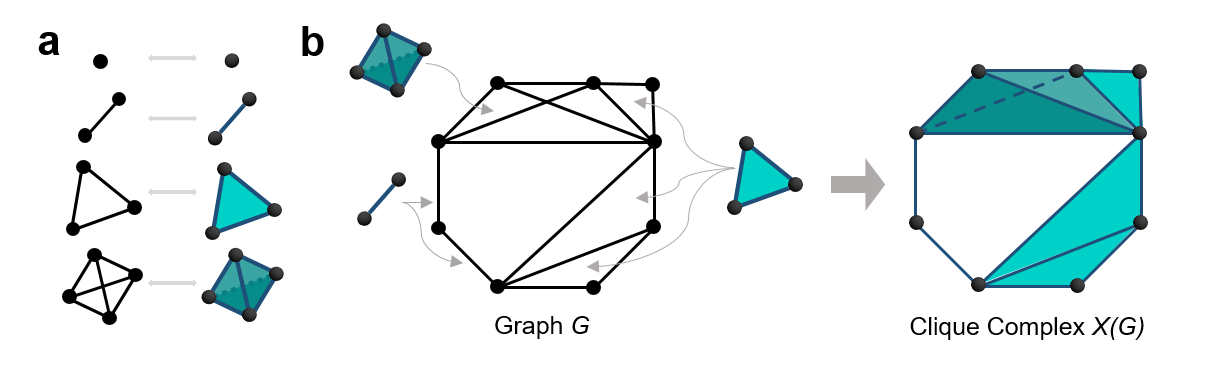}
	\caption{\textbf{From cliques to a clique complex.} \emph{(a)} Cliques are all-to-all connected sets of nodes which we use as "filled in" building blocks. \emph{(b)}The clique complex is created by inserting these building blocks into the completely connected subgraphs of $G$. \label{fig:sfig_ph1}}

\end{figure}

\emph{Cliques} First, we will transform our network of interest into an algebraic object so that we can use powerful computational tools from linear algebra to compute intuitive geometric features. We begin by selecting building blocks from which to assemble larger, mesoscale structures. Drawing on classical graph theory \new{and our intuition about the type of structures we are looking for, we are led to} a natural (and well studied) choice of such blocks: sets of all-to-all connected nodes called \emph{cliques}. In the context of brain networks, cliques are groups of brain regions that are able to rapidly and effectively share information. Formally, a $(k+1)$-clique of a graph $G$ as a set of $(k+1)$ nodes for which all pairwise edges are in G. Thus, a single node is a 1-clique, an edge a 2-clique, a triangle a 3-clique, and so on. Any subgraph of a clique must itself be a clique of lower degree, called a \emph{face}. A \emph{maximal} clique is thus any clique that is not a face. Intuitively, we will think of cliques as "filled in" regions, rather than hollow collections of edges (Fig.~\ref{fig:sfig_ph1}a).

\noindent\emph{Clique Complex} We study the structure formed by all cliques induced by the graph $G$, a \new{combinatorial} object called the \emph{clique complex}\footnote{More specifically, we build the abstract simplicial complex formed from the correspondence of $k$-simplices and $(k+1)$-cliques. See \cite{carlsson2009topology,hatcher2002algebraic,ghrist2014elementary} for more details.} (Fig.~\ref{fig:sfig_ph1}b). The clique complex of a graph $G$ is the collection of all the cliques in $G$, formally denoted $X(G) = \{X_0(G), X_1(G), \dots ,X_N(G)\}$ where $X_k(G)$ is the set of all $(k+1)$-cliques in $G$. Historically, the index is chosen to correspond to the dimension of the enclosed region, and we adopt this index shift here for consistency. The clique complex is an object which allows us to formally manipulate certain important geometric properties (as we explore in more detail in the following sections), and, through these computations, discover  mesoscale features of interest.

\noindent\emph{Chain Group} In order to perform computations, we move from sets of cliques to vector spaces. We define the \emph{chain group} $C_k(X(G))$ (abbreviated to $C_k$ when the underlying clique complex is understood) as the vector space with basis $X_k(G)$. We denote by $\sigma_{i_1, i_2, \dots, i_k} \in C_k(X(G))$ the basis element corresponding to a $(k+1)$-clique on nodes $\{i_0, i_1, \dots i_k\}$. \new{Though this definition can be made for any scalar field, we use vector spaces over the field with two elements, $\mathbb{F}_2 = \{0, 1\}$, as is standard in topological data analysis. Elements of $C_k(X(G))$ are linear combinations of \emph{$k$-chains} which correspond to collections of $(k+1)$-cliques. }

\begin{figure}[h]
	\centering
	\includegraphics[width = 3.5in]{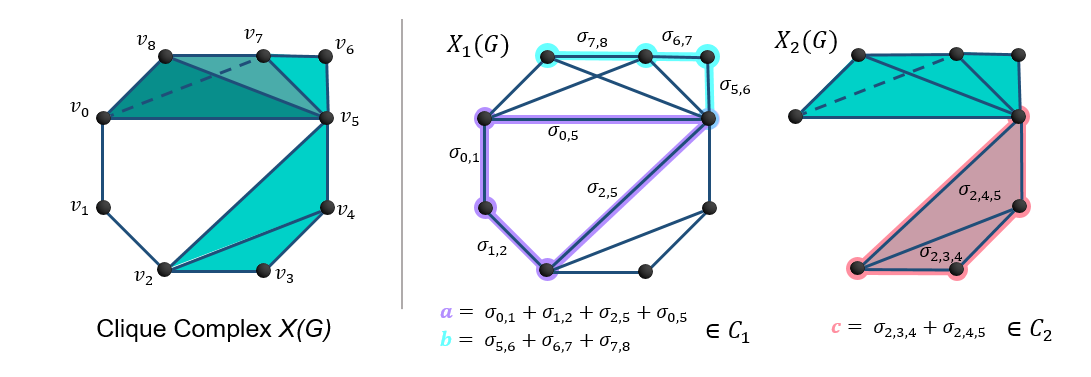}
	\caption{\textbf{Chain group elements are linear combinations of cliques.} See appendix text for a complete description of these graphs. \label{fig:ph2}}

\end{figure}

For example, consider the clique complex $X(G)$ shown in Fig.~\ref{fig:ph2}. Elements of $C_1$ are linear combinations of edges, or $2$-cliques. One such element is $b = \sigma_{5,6} + \sigma_{6,7} + \sigma_{7,8}$, shown in blue in Fig.~\ref{fig:ph2}. This is intuitively an undirected path from $v_5$ to $v_8$ that passes through $v_6$ and $v_7$. We could also take the purple path $a \in C_1$. This path begins at $v_0$ and follows $\sigma_{0,1}$, $\sigma_{1,2}$, $\sigma_{2,5}$, then $\sigma_{0,5}$ which returns us to $v_0$\footnote{Because we work over $\mathbb{F}_2$, this algebraic encoding is not sensitive to clique direction, only the parity of the number of times a clique appears in a chain.} In $C_2$, an element is a linear combination of $3$-cliques. Highlighted in Fig.~\ref{fig:ph2} (right) is one such example: the element $c \in C_2$ with $c = \sigma_{2,3,4} + \sigma_{2,4,5}$. Because we are working in $\mathbb{F}_2$, if we took this path twice, we would have the chain $c+c = \sigma_{2,3,5} + \sigma_{2,4,5} + \sigma_{2,3,5} + \sigma_{2,4,5} = 2\sigma_{2,3,5} + 2\sigma_{2,4,5} = 0$.

\noindent\emph{Boundary Operator} Recall that our goal is to detect topological cavities\footnote{The structure of cycles is subtle and not necessarily indicative of physical cavities in a general sense. However, in the case of these relatively sparse 3D graphs this is usually the case.} in our algebraic object. Cavities exist when cliques are arranged in a loop or shell, but there are no higher dimensional cliques that ``fill in'' the enclosed space -- that is, the shell is not the ``boundary'' of some collection of higher dimensional cliques. To detect this computationally, we use the \emph{boundary operator} $\partial_k : C_k \rightarrow C_{k-1}$, which takes a collection of $(k+1)$-cliques (an element of $C_k$) and sends them to their boundary (an element of $C_{k-1}$). 

\begin{figure*}[h]
	\centering
	\includegraphics[width = \textwidth]{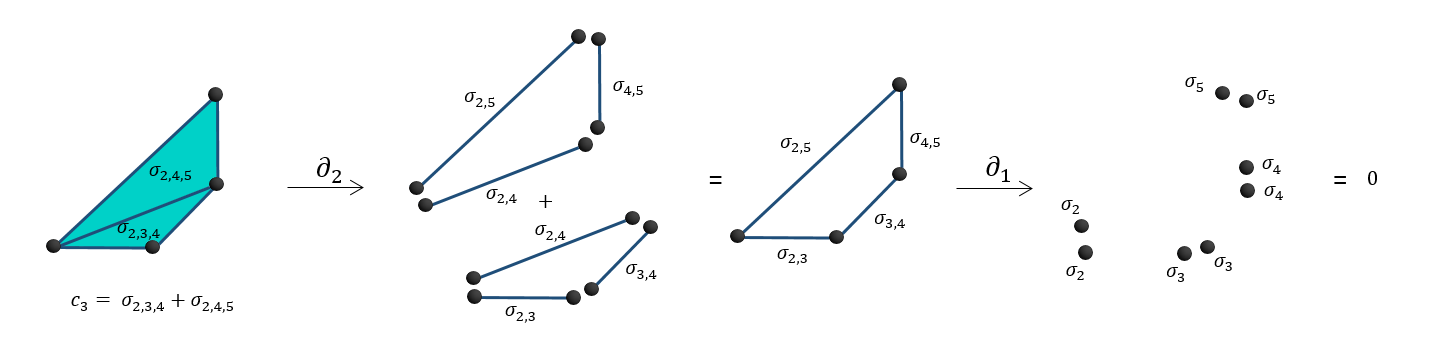}
	\caption{\textbf{Example of the boundary operator in $C_2$.} See appendix text for a complete description of these graphs.
		\label{fig:sfig_ph3}}
	
\end{figure*}

Geometrically, the boundary of a $k$-clique is the family of $(k-1)$-cliques obtained by removing each vertex in succession. The boundary of a contiguous collection of (one or more) $k$-cliques is a ``shell'' of $(k-1)$-cliques surrounding the original collection, inside of which the boundaries of neighboring $(k-1)$-cliques overlap. Detecting this pattern of overlaps computationally is accomplished when chains corresponding to the shared faces cancel. In Fig.~\ref{fig:sfig_ph3} the boundary of $c \in C_2$ is the chain corresponding to the surrounding four edges (2-cliques), as the interior edge ($\sigma_{2,4}$) cancels. Formalizing this intuition, we define the boundary operator (with coefficients in $\mathbb{F}_2$) on the basis $X_k(G)$ to be
$$\partial_k(\sigma_{0,1,\dots,k}) = \sum_{i=0}^{k}\sigma_{0,1,\dots,\hat{i},\dots,k}$$ where $\hat{i}$ indicates that vertex $i$ is not included in the set of vertices that form the clique, and we extend this map linearly to all of $C_k(X(G))$. Then, for example,  in Fig. \ref{fig:sfig_ph3}, $$\partial_2(c_3) = \partial(\sigma_{2,3,4} + \sigma_{2,4,5}) = \partial_2(\sigma_{2,3,4}) + \partial_2(\sigma_{2,4,5})$$
$$= (\sigma_{3,4} + \sigma_{2,4} + \sigma_{2,3}) + (\sigma_{4,5} + \sigma_{2,5} + \sigma_{2,4})$$
$$= \sigma_{3,4} + \sigma_{2,3} + \sigma_{4,5} + \sigma_{2,5}.$$

Because the boundary of $c_3 \in C_2$ is itself an element of $C_1$, we can apply $\partial_1$ to it in turn. As illustrated in Fig.~\ref{fig:sfig_ph3}, $$\partial_1(\partial_2(c_3))= \partial_1(\sigma_{2,3} + \sigma_{3,4} + \sigma_{4,5} + \sigma_{2,5}) $$
$$= \sigma_3 + \sigma_2 + \sigma_4 + \sigma_3 + \sigma_5 + \sigma_4 + \sigma_5 + \sigma_2$$
$$= 2\sigma_2 + 2\sigma_3 + 2\sigma_4 + 2\sigma_5$$
$$ = 0.$$ This example illustrates a crucial property of the boundary operator: $\partial_{k-1} \circ \partial_k = 0$, which will be more thoroughly discussed in the Homology section below.

\noindent\emph{Chain Complex} We now have a boundary operator that lets us move from $k$-chains to $(k-1)$-chains for every $k$\footnote{The boundary of a 0-chain is defined to be 0, since a node is a single point with no geometric boundary.}. These operators link together the chain groups into a sequence
$$ \xrightarrow{\partial_{k+1} = 0} C_k \xrightarrow{\partial_k} C_{k-1} \xrightarrow{\partial_{k-1}} \dots \xrightarrow{\partial_2} C_1 \xrightarrow{\partial_1} C_0 \xrightarrow{\partial_0 =0} 0$$
called the \emph{chain complex} for X(G). This is our fundamental algebraic tool for studying the structure of the network.

\begin{figure*}[h]
	\centering
	\includegraphics[width = \textwidth]{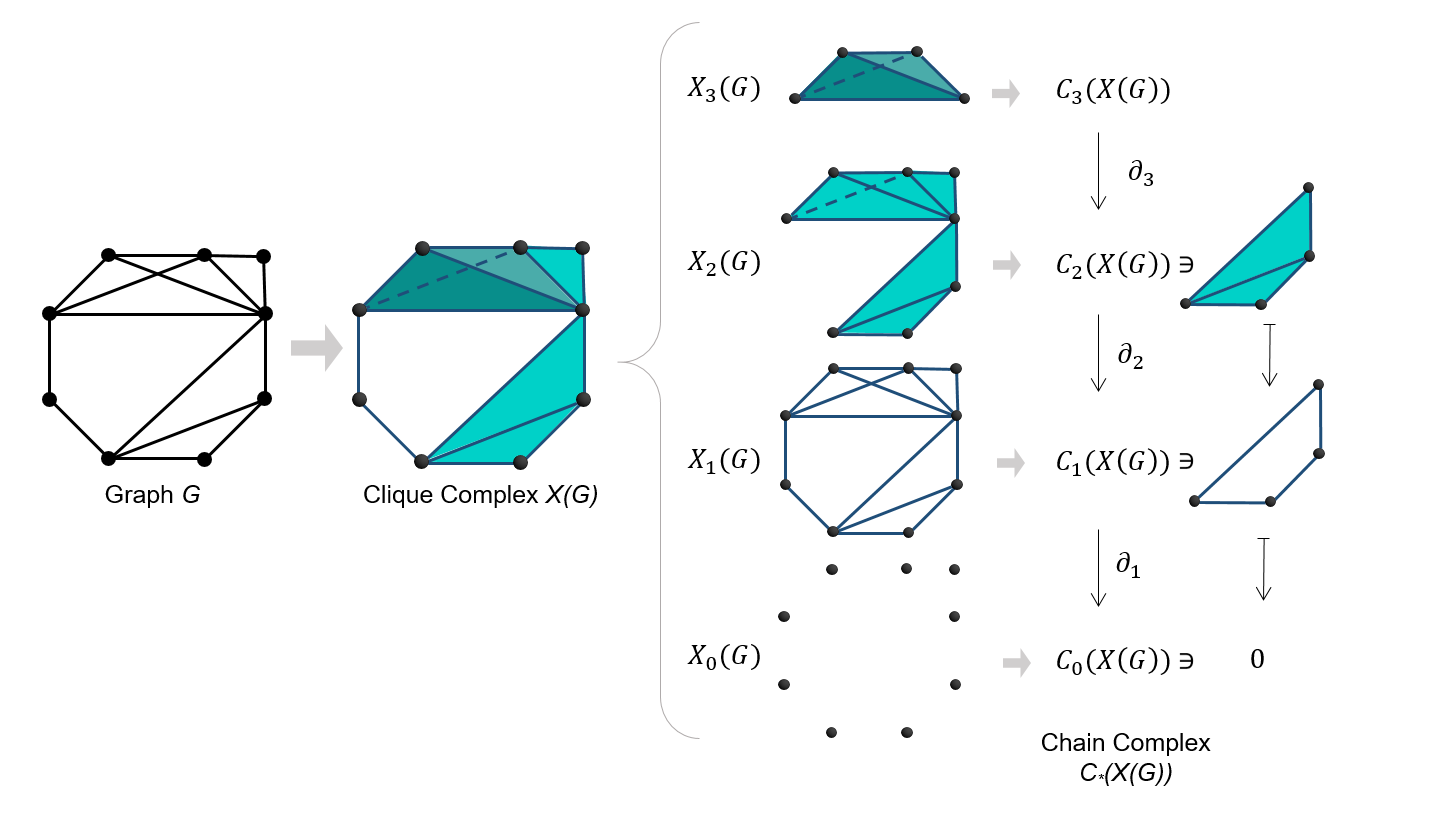}
	\caption{\textbf{Illustration of creating from $G$ the clique complex $X(G)$.} Also shown are the induced chain complex $C_*(X(G))$ and an example of boundary calculations on an element in $C_2(X(G))$.}
	\label{fig:sfig_ph4}
	
\end{figure*}

In summary, we have taken an unweighted, undirected graph $G$ and, from an enumeration of its cliques, formed the clique complex $X(G)$ (Fig.~\ref{fig:sfig_ph4}, left). We then used the cliques of each dimension as basis elements in the chain groups $C_0(X(G)), C_1(X(G)), \dots, C_N(X(G))$ (Fig.~\ref{fig:sfig_ph4}, middle). Finally, we defined the boundary operator $\partial$ that finds the boundary of a chain (which represents a collection of $(k+1)$-cliques), itself a (possibly empty) chain representing a collection of $k$-cliques, and we used this function to string together the chain groups into the chain complex (Fig.~\ref{fig:sfig_ph4}, right).

\subsection{Homology}~

We turn now to the definitions and concepts needed to compute homology. Homology discoveres features of interest in the clique complex by separating \emph{cycles}, mesoscale patterns constructed from cliques, which surround a cavity from those that are the boundary of a collection of cliques.

\begin{figure}[h]
	\centering
	\includegraphics[width = 3.5in]{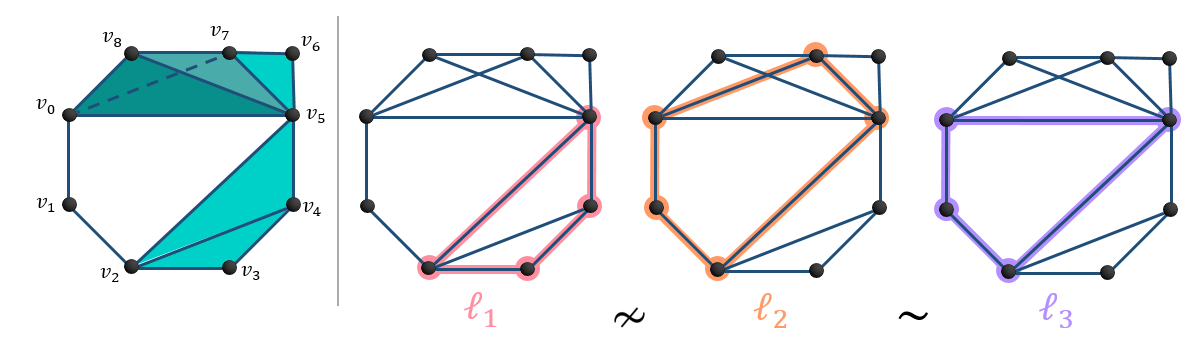}
	\caption{\textbf{Cycles.} Examples of a cycle that is also a boundary ($\ell_1$) and two equivalent, non-boundary cycles ($\ell_2$ and $\ell_3$).}
	\label{fig:sfig_ph5}
	
\end{figure}

\noindent\emph{Cycles} Though we have seen examples of cliques strung together as paths, we are particularly interested in paths that form closed structures called \emph{cycles}, the 1-dimensional analogue of which are graph-theoretic circuits. Consider the three closed circuits in Fig.~\ref{fig:sfig_ph5}, each can be thought of as a linear combination of elements in $C(X_1(G))$. If we begin at any \new{1-clique (node)} on the cycle, for example $\sigma_2$ in $\ell_1$, and traverse each 2-clique in the cycle in order, we will end at our starting \new{1-clique}. Since the boundary of any path $\in C(X_1(G))$ is $\sigma_{end} + \sigma_{begin}$, the boundary of any cycle $\ell \in C(X_1(G))$ must be $$\partial_1(\ell)=\sigma_{end} + \sigma_{begin} = 2\sigma_{begin} = 0$$.

Though we have thus far focused on the familiar notion of cycles built of 2-cliques, the intuitive notion that boundaries should cancel allows us to construct cycles in any dimension. We define a \emph{$k$-cycle} to be any element $\ell\in C_k$ with $\partial_k(\ell) =0$. Since the cycles are exactly the elements that are sent to 0 by the boundary operator, the subspace of $k$-cycles is precisely the kernel (or nullspace), denoted $\text{ker}(\partial_k) \subset C(X_k(G))$.

As noted above, cycles can either surround cavities or a collection of cliques, and since we are strictly interested in cycles of the first type, we must determine how to differentiate between these two options. Fig.~\ref{fig:sfig_ph5} depicts three 1-cycles found in the clique complex shown on the left. Looking strictly at $X_1(G)$, we cannot distinguish which of these three cycles belong to which category.

However if include information about 3-cliques, the separation becomes apparent, in the same way looking at the full depiction of the clique complex in Fig.~\ref{fig:sfig_ph5} (left) makes it apparent that this object surrounds one cavity. We need consider only the image of the boundary map from $\partial_2: C_2(X(G))\to C_1(X(G))$: if a 1-cycle $\ell$ surrounds a collection of higher dimensional cliques, it must in particular surround a collection of 2-cliques (2-faces of these larger cliques). In our example in Fig.~\ref{fig:sfig_ph5}, this means $\ell_1$ is the boundary of some element in $C_2(X(G))$ (this element is $\sigma_{2,3,4} + \sigma_{2,4,5}$). 

We can repeat such an argument for any 1-cycle that surrounds a collection of higher dimensional cliques, which allows us to define \emph{k-boundaries} as elements in $\text{im}(\partial_{k+1}) \subseteq C_K(X(G))$. Furthermore it must be true that $\text{im}(\partial_k) \subseteq \text{ker}(\partial_k)$ per our previous observation that $\partial_k\circ\partial_{k+1} = 0$.

However, not all cycles are necessarily boundaries: $\ell_2$ and $\ell_3$ are in $\text{ker}(\partial_1)$ but neither are elements of $\text{im}(\partial_1)$. The $k$-cycles that surround cavities are thus those that are in $\text{ker}(\partial_k)$ but not $\text{im}(\partial_k)$. However, enumerating cycles in $\text{ker}(\partial_k) - \text{im}(\partial_k)$ is not enough to produce a proper list of cavities in our clique complex, because we will suffer from redundancy. For example, knowing \emph{either} $\ell_2$ or $\ell_3$ tells us the cavity they both enclose exists. Certainly $\ell_2 \neq \ell_3$, but we should consider them equivalent since they both reveal the same feature of our complex. So we need a way to count more carefully.

\noindent\emph{Equivalence} The solution to our enumeration problem will depend on what we regard as "the same". Above we mentioned we should consider $\ell_2$ to be equivalent to $\ell_3$ because they surround the same cavity. How is it that we understood this? We see they both enclose this cavity, while $\ell_2$ also surrounds one 3-clique. But this 3-clique (specifically $\sigma_{0,5,7}$) does not change the cavity or add a new one, so we decided this difference of a higher dimensional clique should be insubstantial, and thus the two cycles are equivalent. Generalizing this example provides a method for correctly enumerating the cavities in the complex.

Two $k$-cycles, $\ell_i$ and $\ell_j$, are called \emph{equivalent} if their sum, (working over $\mathbb{F}_2$) $\ell_i + \ell_j$ is the boundary of a $(k+1)$-chain, e.g. $\ell_i \sim \ell_j$ if $\ell_i + \ell_j \in \text{im}(\partial_{k+1})$. In Fig.~\ref{fig:sfig_ph5}, we have 
\begin{eqnarray*}
	\ell_2 + \ell_3&=& (\sigma_{2,5}+\sigma_{5,7} + \sigma_{7,0} + \sigma_{0,1} + \sigma_{1,2})\\
	&&\;\; + (\sigma_{2,5} + \sigma_{5,0} + \sigma_{0,1}+ \sigma_{1,2})\\
	&=& \sigma_{5,7} + \sigma_{7,0} + \sigma_{0,5}\\
	&=&\partial_2(\sigma_{0,5,7}) \in \text{im}(\partial_2)
\end{eqnarray*}
so indeed we see $\ell_2 \sim \ell_3$.

This, finally, provides us with a proper count: if we only count one cycle from each set of (non-trivial) equivalent cycles, then we will have precisely the number of topological cavities of a given dimension within the clique complex. The clique complex in Fig.~\ref{fig:sfig_ph4} by eye has only one cavity surrounded by 1-cycles, and our computations to come to the same conclusion. Notice that any closed loop of 2-cliques either is equivalent to $\ell_2$ or it is strictly a boundary of higher dimensional cliques and thus is trivial. So, as desired, we have a sole 2-dimensional cavity. 

The \emph{equivalence class} of a $k$-cycle $\ell$ is $[\ell] = \{\nu \in Z_k|\nu \sim \ell\}$. Note the equivalence class of boundary loops $b \in \text{im}(\partial_k)$ contain the empty set, since $b - \emptyset  = b \in \text{im}(\partial_k)$. This means for any $\ell \in \text{ker}(\partial_k)$ and $b \in \text{im}(\partial_k)$, we have $\ell + b \sim \ell + \emptyset \sim \ell$, confirming our requirement that cycles differing by boundaries are equivalent. By abuse, it is common to refer to an equivalence class of $k$-cycles as a $k$-cycle, and we will continue with this convention.

\noindent\emph{Homology Groups} The heavy lifting is now complete and we are left with only the formal definition of homology to conclude the section. 
Recalling the equivalence classes we have discussed above, we define the homology group of dimension $n$ as 
$$ H_n := \ker(\partial_n)/\text{im}(\partial_{n+1})$$
which is simply the vector space spanned by \new{equivalence classes of} $n$-cycles. The dimension of $H_n$ is the number of nontrivial $n$-cycles and thus the number of $(n+1)$-dimensional topological cavities of our clique complex. In summary we can now take a graph of nodes and edges, convert it to an algebraic object called the clique complex, then use the boundary operator to find equivalence classes of cycles that describe essential mesoscale architecture of our network in the form of topological cavities.

\subsection{Homology for Weighted Networks: Persistent Homology}
While homology detects cavities in binary graphs, the DSI data (and many other sources in biology) create a weighted network. \emph{Persistent homology} was developed \cite{carlsson2009topology,zomorodian2005computing} to address this situation. The method uses the edge weights to unravel the weighted network into a sequence of binary networks on which we can then compute homology, in a manner related to but more principled than standard thresholding techniques. Overall persistent homology perceives how the features seen with homology evolve with the weighted network.

\begin{figure}[h]
	\centering
	\includegraphics[width = \textwidth]{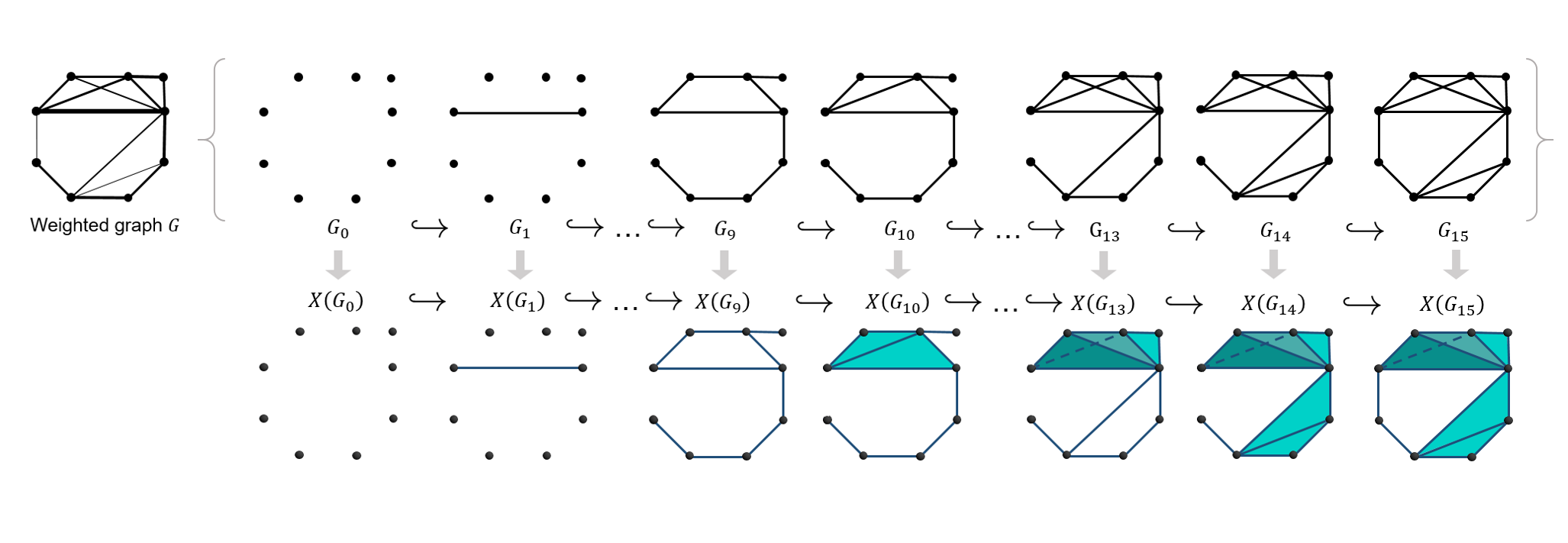}
	\caption{\textbf{Filtrations and inclusion maps.} Edge weights indicated by line thickness induce a filtration on the weighted graph $G$. The inclusion maps $G_i \hookrightarrow G_{i+1}$ induce inclusion maps on the corresponding clique complexes $X(G_i) \hookrightarrow X(G_{i+1})$.\label{fig:sfig_ph6}}

\end{figure}

\noindent\emph{Filtrations} Given $G$ a weighted network, we first construct a sequence of binary graphs that will allow us to use homology on each graph in the sequence. The edge weights induce a natural ordering on the edges from highest to lowest weight. Then, beginning with the empty graph, we replace edges following this ordering. This process creates a \emph{filtration} $$G_0 \subset G_1 \subset \dots \subset G_{|E|} = G$$ where each $G_{i+1}$ contains one more edge than $G_i$. Since $G_{i+1}$ contains $G_i$ (and one more edge), we obtain an inclusion map $i:G_i \hookrightarrow G_{i+1}$ which describes how $G_i$ maps into $G_{i+1}$. In our case this is quite natural, $G_i$ is sent to itself, now a subgraph of $G_{i+1}$ (Fig.~\ref{fig:sfig_ph6}, top row).

Having an inclusion of $G_i$ into $G_{i+1}$ means we can also get an inclusion of $X(G_i)$ into $X(G_{i+1})$ in a similar fashion, where cliques in $X(G_i)$ map to their corresponding selves in $X(G_{i+1})$ (Fig.~\ref{fig:sfig_ph6}, bottom row).

\begin{figure}[]
	\centering
	\includegraphics[width = 3in]{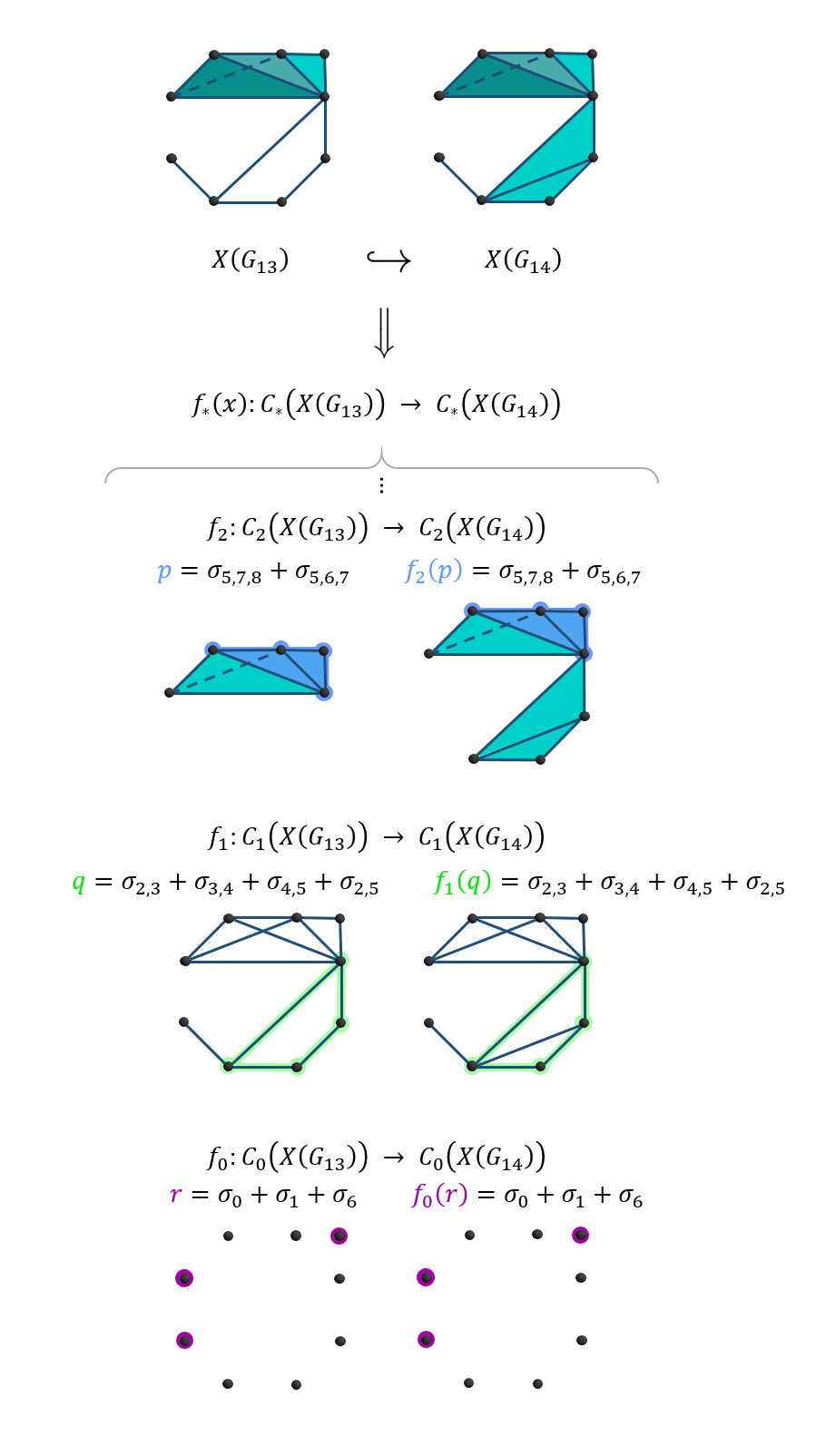}
	\caption{\textbf{Inclusion maps between clique complexes induce maps between the corresponding chain complexes.} See appendix text for a complete description of these graphs. \label{fig:sfig_ph6_5}}
	
\end{figure}

But now knowing how one clique \new{complex} maps into the next clique complex means we get maps between the chain groups as well. For example, in Fig.~\ref{fig:sfig_ph6_5} we look only at the inclusion of $X(G_{13})$ into $X(G_{14})$. This inclusion map tells us how to take cliques from $X(G_{13})$ and fit them into $X(G_{14})$, which means we can figure out how to take some combination of cliques and fit them into $X(G_{14})$ as well. The functions that perform this task are defined 
$$f_*:C_*(X(G_{13})) \rightarrow C_*(X(G_{14})) $$
where the $*$ refers to the set of functions indexed by dimension. We show the first three with examples in Fig.~\ref{fig:sfig_ph6_5}. If we have a 0-chain $r = \sigma_0 + \sigma_1 + \sigma_6 \in C_0(X(G_{13}))$, it gets mapped by $f_0$ to a chain in $C_0(X(G_{14}))$, explicitly $f_0(r) = \sigma_0 + \sigma_1 + \sigma_6$. 

We can do this in the higher dimensions as well. Figure \ref{fig:sfig_ph6_5} also shows the green 1-chain $q = \sigma_{2,3} + \sigma_{3,4} + \sigma_{4,5} + \sigma_{2,5} \in C_1(X(G_{13}))$ and how it maps into $C_1(X(G_{14}))$ as well. It is interesting here to note that in $C_1(X(G_{13}))$, the 1-chain $q$ is also a 1-cycle, but this is not the case in $C_1(X(G_{14}))$. Again we can move to the 2-chains and observe how $p = \sigma_{5,7,8}+ \sigma_{5,6,7}$ is sent to $f_2(p) = \sigma_{5,7,8} + \sigma_{5,6,7} \in C_2(X(G_{14}))$.

Generally filtrations are a powerful way to understand weighted networks. Here, we will use these chain maps $f_*$ to track particular chains throughout the filtration to see how they may change as new edges (and thus cliques) are added.

\noindent\emph{Persistent Homology} As we are interested in cycles, we now turn to tracking specifically cycles throughout the filtration. A $k$-loop is a $k$-chain, so it can be tracked horizontally from clique complex to clique complex in the filtration. Additionally, we have vertical boundary maps that tell us if the $k$-loop in question is a cycle or a boundary loop within the particular clique complex. More generally we are combining the information from the filtration and its between-complex induced maps (Fig.~\ref{fig:sfig_ph6},~\ref{fig:sfig_ph6_5}) with the boundary loop information from the within-complex boundary operators (Fig.~\ref{fig:sfig_ph4}) to observe how cycles change as we add edges of decreasing weight.      

\begin{figure}[h]
	\centering
	\includegraphics[width = \textwidth]{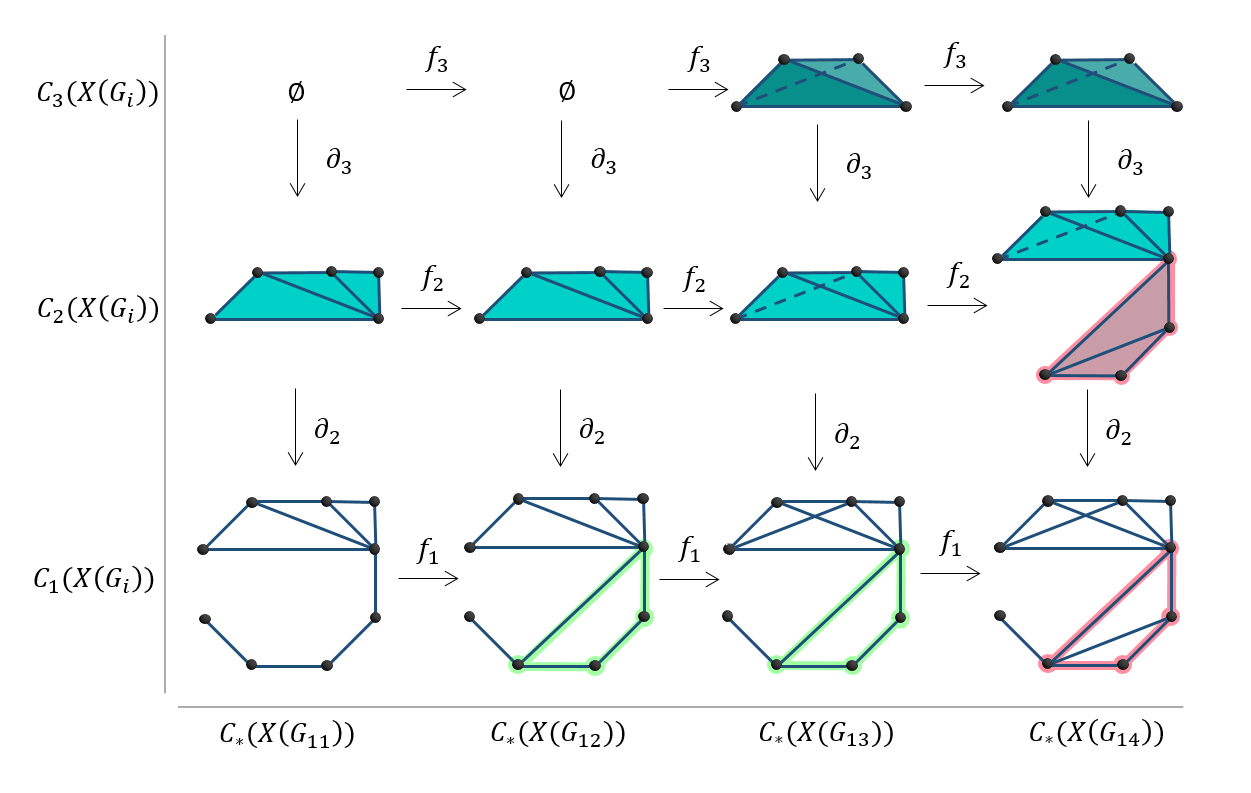}
	\caption{\textbf{Illustration of the persistence complex of the weighted graph $G$.} The green 1-cycle is first seen in $X(G_{12})$, is mapped through filtrations, and finally becomes the boundary of a collection of 3-cliques (pink) in $X(G_{14})$.}
	\label{fig:sfig_ph7}
	
\end{figure}

Formally these maps and complexes form the \emph{persistence complex} of our weighted graph $G$ (Fig. \ref{fig:sfig_ph7}).
Armed with inclusion and boundary maps between chain groups, we can compute the homology of each graph in the filtration and therefore obtain maps $H_*(X(G_i)) \rightarrow H_*(X(G_{i+1}))$ that describe how cycles (equivalence classes of cycles) in $X(G_i)$ change (map directly, shrink in length, become a boundary loop) in $X(G_{i+1})$.

For example, in Fig.~\ref{fig:sfig_ph7} we see the green 1-cycle first appears in $G_{12}$. We say the cycle is \emph{born} at this edge density $\rho_{birth} = (\text{\# edges present})/(\text{\# edges possible}) = 12/36$. The green cycle continues to exist until it maps to a cycle that is the boundary of the pink $2$-chain in $C_2(X(G_{14}))$. Since this cycle is now a boundary, it is equivalent to the trivial cycle in $H_1(X(G_{14}))$. We say the cycle \emph{dies} at this edge density $\rho_{death} = 14/36$.

\begin{figure}
	\centering
	\includegraphics[width = 2.5in]{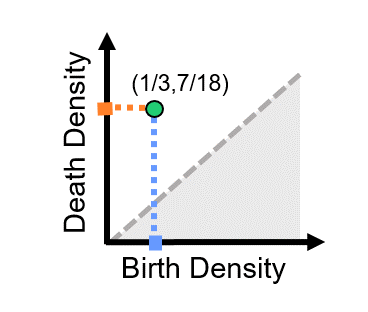}
	\caption{\textbf{Example persistence diagram for green cycle shown in Fig.~\ref{fig:sfig_ph7}.} See appendix text for a complete description of these graphs. }
	\label{fig:sfig_ph8}
\end{figure}

Cycles that exist over many edge additions must evade becoming triangulated by cliques, thus becoming a boundary. Therefore we consider such cycles more essential if they \emph{persist} for many edge additions. We measure cycle persistence in two ways. First we record cycle \emph{lifetime} $l = \rho_{death} - \rho_{birth}$, which is commonly used in persistent homology calculations \cite{carlsson2009topology} and displayed on a persistence diagram. For our cycle which is born at $\rho = 12/13 = 1/3$ and dies at $\rho = 14/36 = 7/18$, we see an example persistence diagram in Fig.~\ref{fig:sfig_ph8}. However, recent work \cite{bobrowski2015maximally} suggests alternatively considering $\pi = \rho_{death}/\rho_{birth}$ which allows for cycle persistence comparison at difference scales and underscores the importance of cycles forming at low edge densities.

To summarize, persistent homology tracks interesting connection patterns (cycles) through network frames induced by edge weights, recovering a parameter-free perspective on essential structural features in a weighted network.

\subsection{Comparison with alternative loop-finding algorithms}~
One may ask how our method compares with other loop-finding algorithms. While such programs can be powerful, two fundamental differences exist. The first is in the definition of cycles identified. Recall that we extract equivalence classes of cycles, so we will find only cycles that enclose a structural cavity, while loop-finding algorithms will extract all loops that are boundaries of higher cliques \cite{tucker2006chapter}. Additionally, persistent homology detects cycles in multiple dimensions with much less computational effort than loop algorithms \cite{johnson1975finding}.

Additionally one might ask how small changes in edge weights or edge ordering may affect these findings. Cohen-Steiner et al. showed generally small changes in the edge ordering will result in small changes in the persistence diagram \cite{cohen2007stability}. This makes persistent homology relatively robust to noise and consequentially a powerful tool in neuroscience \cite{giusti2016two}.

\section{Cycles in the Average DSI Data}

\new{To understand the function non-boundary cycles may have in the structural brain network, we recover the minimal generator at $\rho_{birth}$ for each persistent homology class found in the averaged DSI data (Fig.~\ref{fig:fig4}c). These cycles for all 20 of the 2D cavities and the two 3D cavities are shown below in Fig.~\ref{fig:allcycles1}, \ref{fig:allcycles2}, respectively. To summarize this information we plot all minimal representatives with edges weighted by their participation in minimal representatives. This summarization is similar to the frequency scaffold \cite{lord2016insights,petri2014homological} in Fig.~\ref{fig:scaffold}, though here we are unable to assign one minimal representative to each persistent equivalence class so if an edge is part of any of the minimal representatives of one equivalence class it gets an added weight of one. Cycles reach most areas of the brain, and as seen in Fig.~\ref{fig:allcycles1}, many follow the cortical to subcortical theme. The edge involved in the highest number of dimension-one minimal generators in the average DSI data links the left and right thalamus. For dimension 2 we see each edge only exists within one minimal generator.}
	
\begin{figure}
	\centering
	\includegraphics[width = 4.5in]{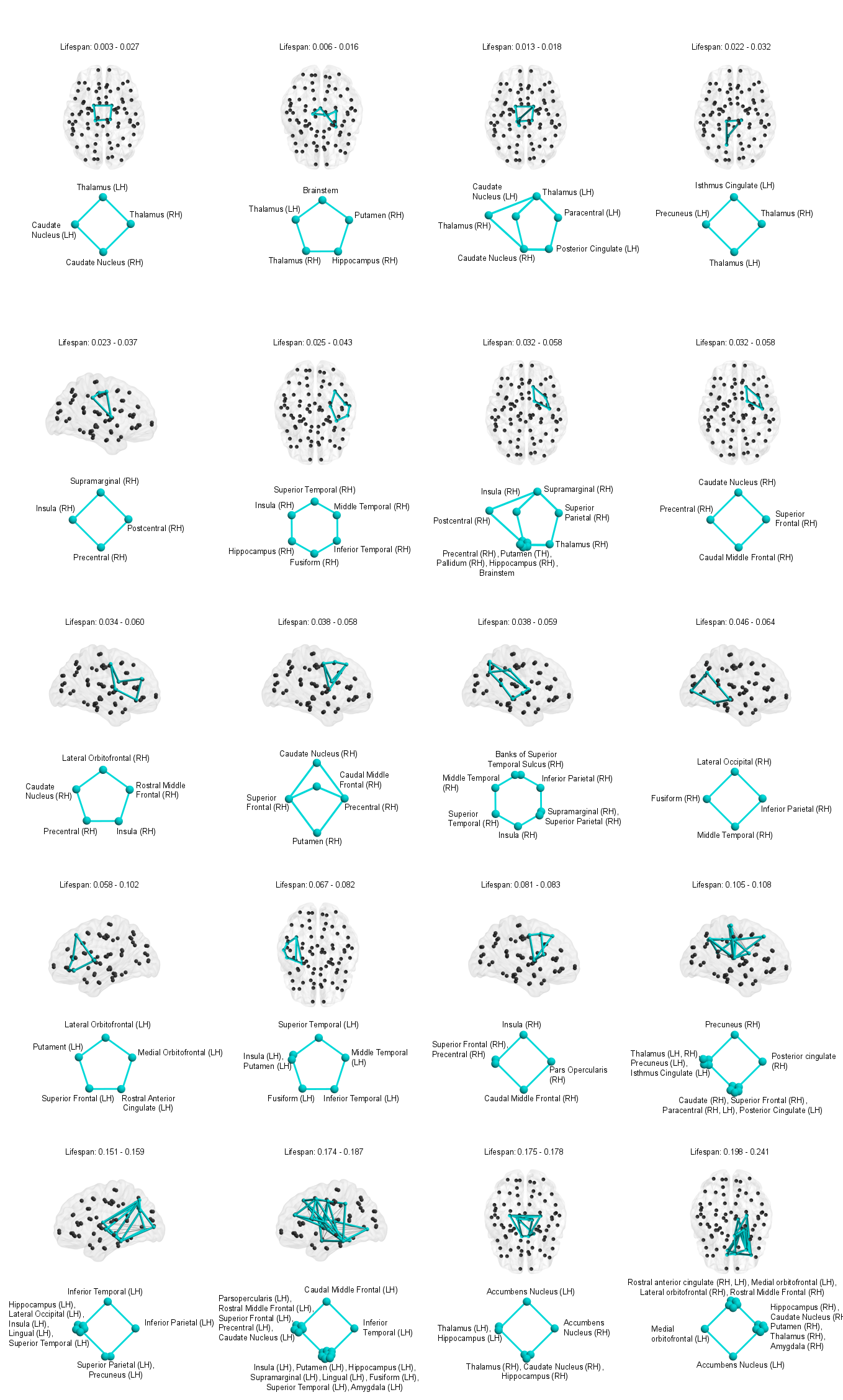}
	\caption{\new{Minimal representatives at $\rho_{birth}$ of all 2D cavities found in the average DSI data, listed in order of increasing birth density. For each topological cavity, the lifespan ($\rho_{birth}$ - $\rho_{death}$), location in the brain, and schematic is shown. For the third, seventh, and tenth appearing cavities, we could not isolate a single equivalence class. }}
\end{figure}

\begin{figure}
	\centering
	\includegraphics[width = 4in]{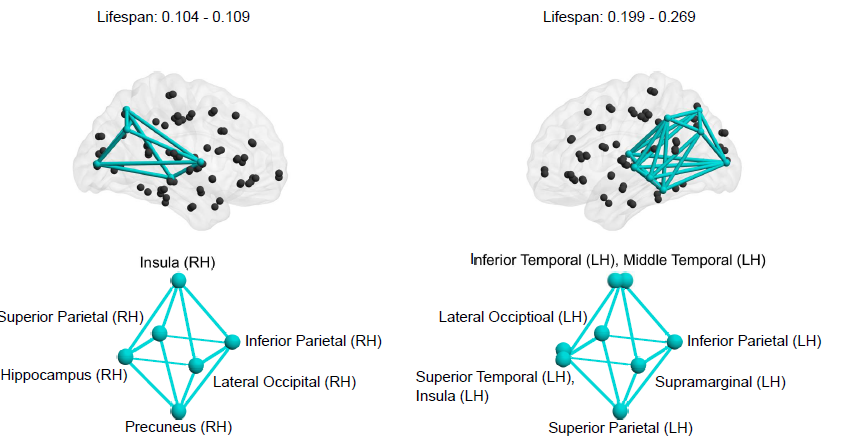}
	\caption{\new{Minimal representatives at $\rho_{birth}$ of all 3D cavities found in the average DSI data, listed in order of increasing birth density. For each, the lifespan ($\rho_{birth}$ - $\rho_{death}$), location in the brain, and schematic is shown.}
		\label{fig:allcycles2}}
\end{figure} 

\begin{figure}
	\centering
	\includegraphics[width = 4in]{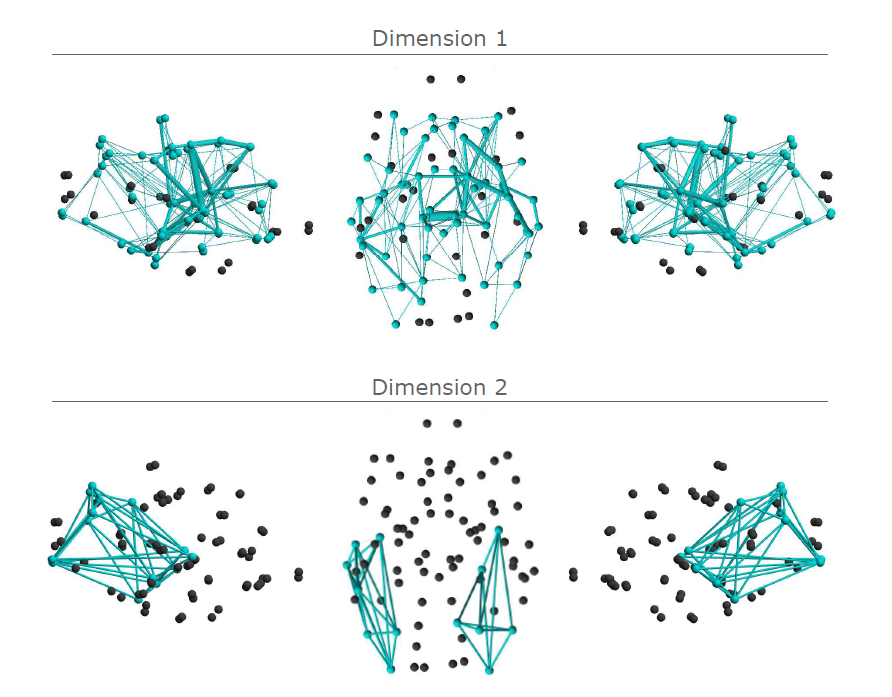}
	\caption{\new{Spatial distribution of minimal generators at $\rho_{birth}$ of 2D (top) and 3D (bottom) persistent cavities. Edges sized by number of minimal generators in which they participate.}}
	\label{fig:scaffold}
\end{figure}

\section{Cycles in Individuals}

Though we detected \new{persistent cavities} in the group-averaged DSI network using persistent homology, we also ask whether these cyclical patterns of connectivity and the corresponding cavities exist in multiple individuals and in multiple scans acquired from the same individual. To address this question, \new{we asked whether a similar geometric loop is seen and whether a similar topological cavity is present in each scan.}

\subsection{Considerations in per scan cycle validation}

Persistent homology is a powerful tool with which to understand the mesoscale homological features of a weighted network. Determining all minimal generators for \new{each of the long-lived topological cavities} gives a finer resolution of such features which can have biological implications, as is the case with our DSI data. Isolating \new{all minimal generators for each homology class} additionally gives a geometric interpretation to these \new{cavities}. Then each \new{cavity} can be viewed from a biological, topological, and to a lesser extent geometric perspective.

This presents a challenge when looking for the ``same'' \new{persistent homology classes} in another clique complex. From the neuroscience perspective, two \new{minimal} cycles may be similar if the cycles include the same brain regions, or if the group of regions forming the cycle performs the same function as those in the \new{first}. Geometrically we would perhaps require the same rigid shape of two cycles to call them similar. Finally through the lens of topology we might call \new{two minimal cycles} in two different complexes similar if we can find a map between the complexes which takes one cycle to the other. \new{This is quite complicated so perhaps we could instead ask if the minimal cycle of a homology class in the first clique complex exists in the second as a cycle in a nontrivial homology class but not necessarily as the minimal generator.} This is the least stringent and is an area of active research \cite{carlsson2010zigzag,dey2014computing}. Yet persistent homology is fundamentally a tool for understanding topological features, so it is important that we weight \new{the topological considerations and ideas of similarity} heavily when we compare cycles found in the average data \emph{versus} individual scans.

With these three perspectives in mind, we present the set of rules used in this paper to define whether a \new{persistent homology class} found in an individual scan was the ``same'' as the \new{persistent homology class} in the average network. Though \new{convoluted}, we argue that with the tools available, these requirements for similarity adequately capture \new{some flexibility of} topological similarity while being conservative enough to generally preserve the biological function of the cycle as well. 

\new{We consider each persistent homology class in turn. For a given persistent homology class found in the average DSI connectome, we denote the set of minimal generators of the homology class at $\rho_{birth}$ by $L$ with elements $\ell_i$ for $i = 0,1,2,...m$. Then for each $\ell_i$ there is a set of nodes $N_i$ containing the nodes within this representative. We require both a loop formed by connections between at least one of $N_0, N_1, \dots, N_m$ and a similar topological cavity to exist.}

\begin{enumerate}
	\item \new{Nodes connected in a loop. We first take the subgraph on $N_i$ and ask if there is precisely one non-trivial homology class at any edge density. We then show the connection pattern at the edge density at which this class first appears, for example in Fig.~\ref{fig:sfig_1cycles}d we find a non-trivial cycle equivalence class in every scan, though this is not seen in panel \emph{(h)}. This first allows us to ask if these nodes ever form a non-trivial cycle throughout the filtration, which is possibly of interest from a geometric and neuroscience perspective. We also use this first test as a filter to see in which scans could these nodes surround a topological cavity. Then if we find a non-trivial cycle formed by any of $N_0, N_1, \dots, N_m$, this scan passes to the next stage.}

 \item \new{Similar topological cavity. We then ask if a similar topological cavity exists. The algorithm from \cite{henselmannovel} returns the birth density (and thus birth edge) of each persistent homology class. In order of increasing birth density, we ask if any of the nodes in $N_0, N_1, \dots, N_m$ are in the birth edge. If this is true, we call this a similar cavity in an individual scan if any of the following hold:}
 
	 \begin{enumerate}
	 	\item \new{Let $m_0, \dots, m_k$ be minimal generators of this homology class in the individual scan at $\rho_{birth}$. If any of $m_0, \dots, m_k$ are the same as one of $\ell_0, \dots, \ell_M$ or are in the same equivalence class, then we call this a similar topological cavity and we are done. This is the most straightforward and was most frequently observed within the unnormalized data. See Fig.~\ref{fig:sfig_case3}a for an example.}
	 	
	 	\item \new{If there is some cycle within this non-trivial homology class at $\rho_{birth}$ formed by at least all but one node of some $n_i$, along with no more than two additional nodes, and nodes from $n_i$ are in the original order along the cycle, we call this similar. See Fig.~\ref{fig:sfig_case3}b for an example.} 
	 	
	 	\item \new{If either \emph{(a)} or \emph{(b)} hold at any $\rho$ with $\rho_{birth} \leq \rho < \rho_{death}$, we call this a similar topological cavity. See Fig.~\ref{fig:sfig_case3}c for an example where the cycle is validated at $\rho > \rho_{birth}$. At $\rho_{birth}$, a minimal cycle contains seven nodes, four of which are the thalamus and caudate nucleus from both hemispheres. Following the minimal cycles throughout the lifetime of this persistent cavity we find at some edge density before $\rho_{death}$, a minimal representative consists of exclusively the thalamus and caudate nucleus regions from both hemispheres.}
	 
	 \end{enumerate}

\begin{figure}[h!]
	\centering
	\includegraphics[width = 3.5in]{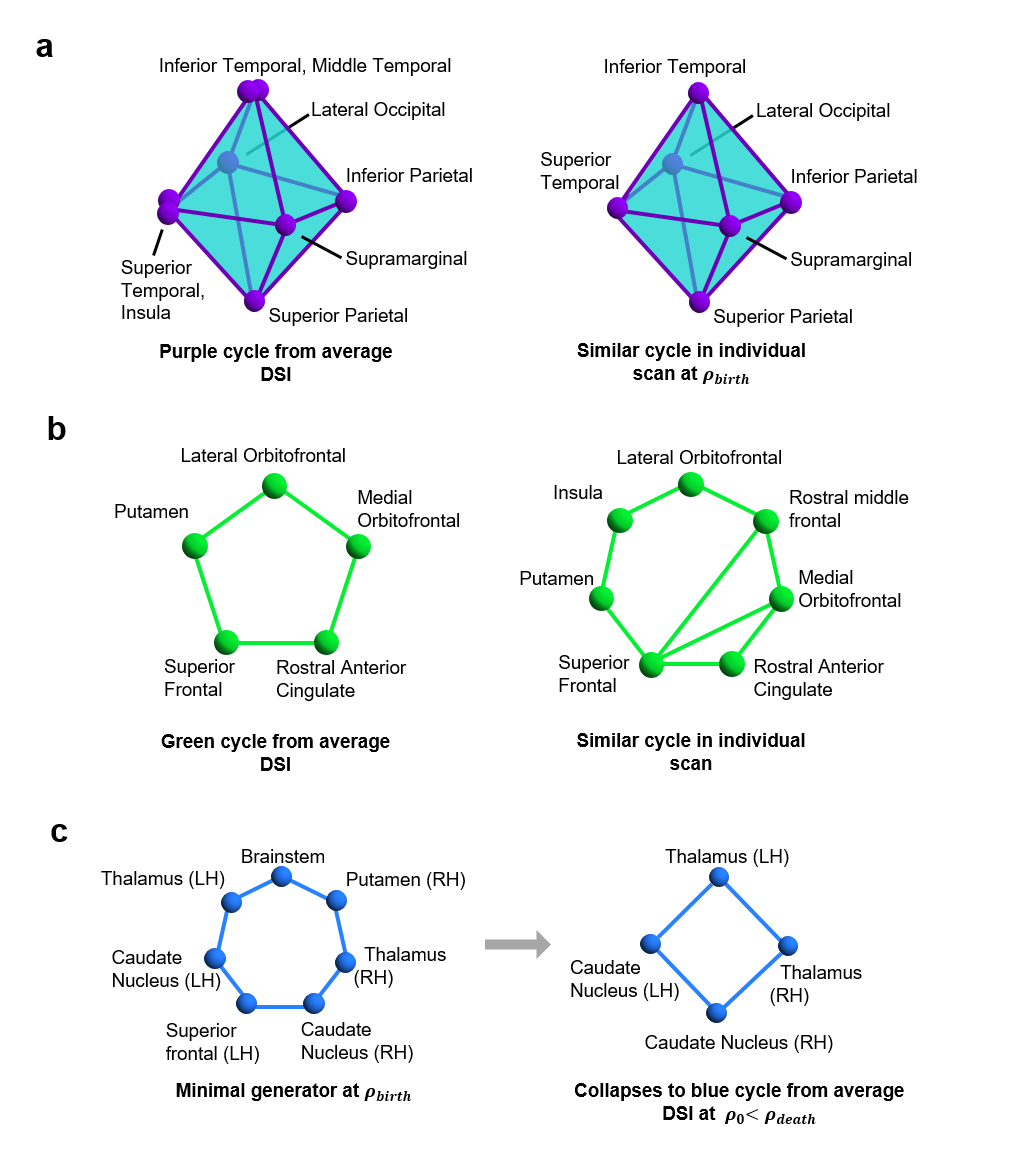}
	\caption{\new{\textbf{Determining similar topological features in individual scans} \emph{(a)} Example of an individual scan in the unnormalized dataset in which we see one of the minimal generators of the purple cycle from the average DSI data (left) at $\rho_{birth}$ for a persistent homology class (right, minimal generator shown). \emph{(b)} Cycles from brains in individual scans may be similar in shape to those seen in the average DSI network. Example of a cycle we call similar (right) to the green minimal cycle in the average network (left). The insula and rostral middle frontal regions in the cycle on the right are needed here for the minimal cycle to exist, but a slightly larger loop contains all of the original nodes in the same order. Finally this cycle still connects subcortical to cortical regions. This example was found in the unnormalized dataset, shown at $\rho_{birth} = 0.08$. \emph{(c)} Example of a minimal generator that begins as a large loop but collapses onto our original cyclic connection pattern at some edge density $\rho_0$, $\rho_{birth} < \rho_0 < \rho_{death}$. This example is from an individual scan in the unnormalized dataset with $\rho_{birth} = 0.008$, $\rho_0 =0.022$, and $\rho_{death} = 0.036$}.
		\label{fig:sfig_case3}}
	
\end{figure}

\new{The first test covers the possibility of the same biological and geometric feature occurring in the individual scan. The second is perhaps the most important, however, because it allows for matching the topological cavity itself. It is important to remember the topological cavities are the features of interest, not the precise \new{cycles} themselves, though the two are clearly intertwined. With the focus on the topological holes, the rationale for the three subrules 2a, 2b, and 2c, is more clear. Though labor intensive, this lets us keep the topological perspective when determining cycle similarity. Moreover, the rationale for focusing on cavities and not specific connections is similar to why large-scale organization such as communities \cite{betzel2016modular}, cores \cite{hagmann2008mapping}, and rich-club organization \cite{van2011rich} are studies with increased intensity. Composed of a plurality of interacting brain regions, these types of structures, and not the individual brain regions nor connections, form computational units that theoretically act to help segregate and integrate information flow across the brain.}.

\new{One clear drawback of this method is the possibility of false negatives. For example, a persistent homology class may have been born which is similar to the cycle in the average data, yet the beginning edge did not include any of the cycle nodes and thus we would not detect this following the above procedure. This is a first attempt to identify similar topological cavities across subjects, and we expect more robust algorithms to be a topic of future research.}

\end{enumerate}

\subsection{Confirming topological cavities in contralateral hemisphere}

\begin{figure}[h]
	\centering
	\includegraphics[width = \textwidth]{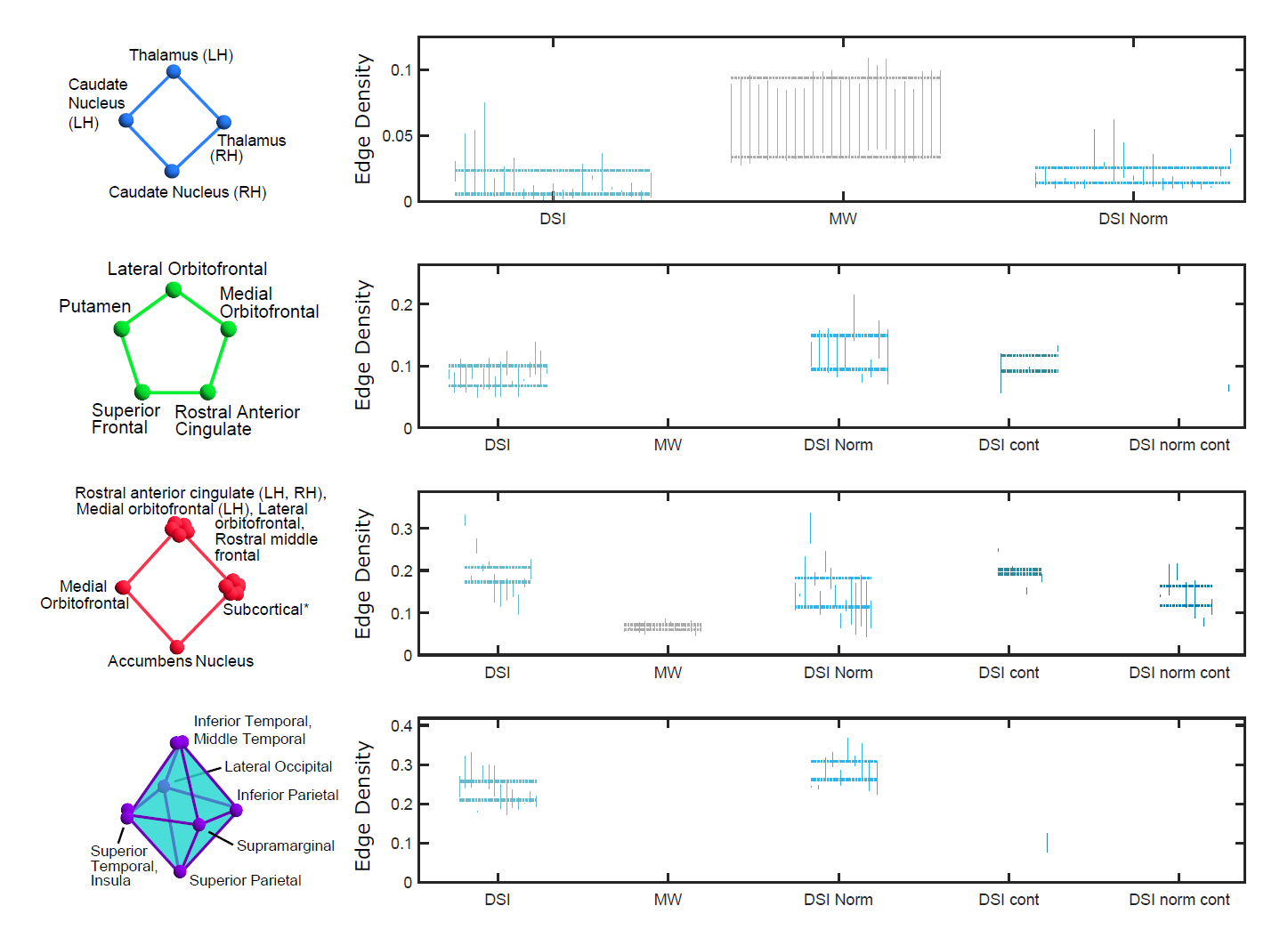}
	\caption{\new{Validation of similar topological cavities in additional data. For each of the four minimal generators highlighted in the main text, bars indicating cavity lifetime for all collected data. Dotted lines indicate average birth or death edge density.}}
	\label{fig:sfig_main4}
	
\end{figure}

In the main text we show validation of the four highlighted cycles in individual scans. Following the procedure above, we next ask if these cycles are seen in the contralateral hemisphere to asses symmetry of these features. Figure \ref{fig:sfig_main4} shows these features are seen in the contralateral hemisphere, though with less frequency than in the original.

\subsection{Cavities in the normalized dataset}

\new{When studying the network formed from DSI, it is important to consider any potential bias created by the different sizes of the 83 brain regions. To account for this potential bias, we normalized the original network of streamline counts by the geometric mean of the end point region sizes and checked to see which cycles were still present \cite{hagmann2008mapping}. More precisely, the normalized edge weight $A_{i,j}$ between nodes $i$ and $j$ is $\text{streamline count}_{ij}/(\text{volume}_i \text{volume}_j)^{1/2}$ \cite{bassett2011conserved}. }

\new{After this normalization, we asked if the cycles found in the streamline counts data are present in the normalized networks. Figs.~\ref{fig:sfig_main4} (DSI Norm, DSI Norm cont) show the cycles are found to a similar extent across scans in the original and contralateral hemispheres.}

\subsection{Locating all cavities from the group-averaged DSI in the minimally wired networks}

\begin{figure}[h]
	\includegraphics[width = \textwidth]{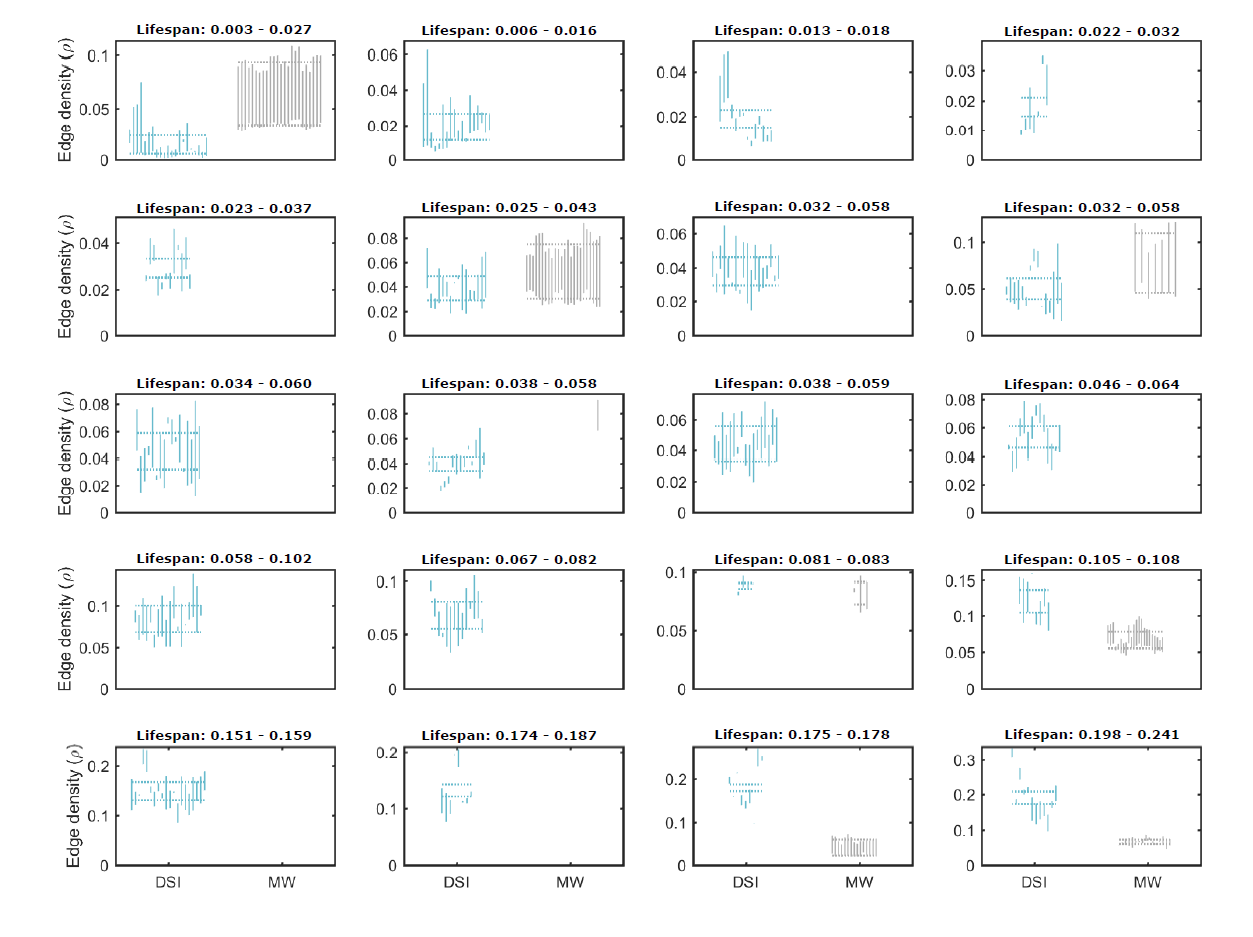}
	\caption{\new{Lifetimes of all 20 persistent 2D cavities in the individuals (black bars) and minimally wired models (gray bars). Dashed lines indicate the average birth and death densities of each class.}}
	\label{fig:sfig_lifetimes}
\end{figure}

\new{Noting many persistent cycles seem likely sampled from the minimally wired distribution of persistent cycles, we asked if we detect the 20 cycles observed in the average data in the null model. Figure \ref{fig:sfig_lifetimes} show the lifespan of each of these persistent cycles within the individual scans (black) and the minimally wired null model (gray). Each vertical bar represents a persistent cavity within a scan, and scans where the cavity was not validated are removed. Average birth and death densities are indicated with horizontal dashed lines. We surprisingly see very few of the persistent homology classes of the DSI data have counterparts in the minimally wired null model. Of those that do, often the average birth and death times are quite different, underscoring the importance of the filtration in this method.}

\begin{figure}[h]
	\centering
	\includegraphics[width = 4in]{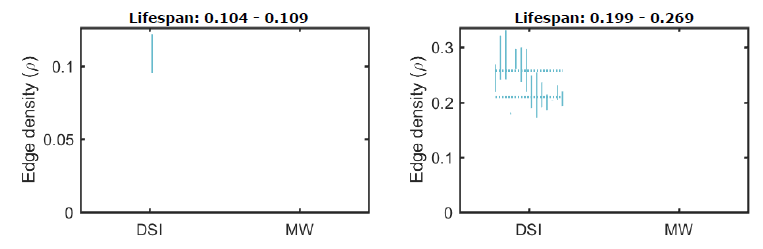}
	\caption{\new{Lifetimes of both persistent 3D cavities in the individuals (black bars) and minimally wired models (gray bars). Dashed lines indicate the average birth and death densities of each class.}}
	\label{fig:sfig_lifetimes2}
\end{figure}

\subsection{Cortical cavities}

\begin{figure}[h!]
	\centering
	\includegraphics[width = 5.5in]{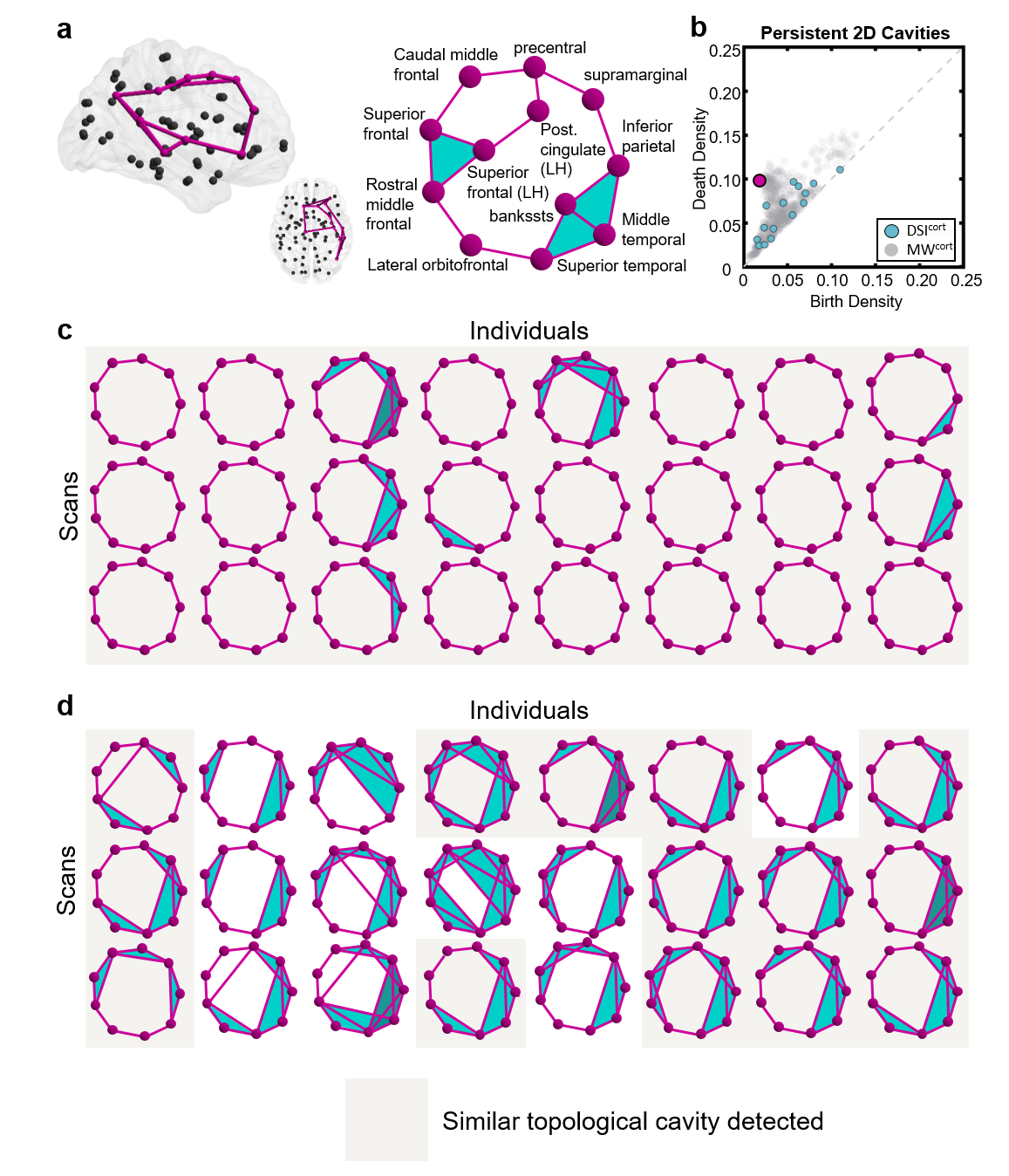}
	\caption{\textbf{Recovered 1-cycle on nine nodes.} \emph{(a)} \new{Minimal representatives at $\rho_{birth}$} shown in the brain (\emph{left}) and as a schematic (\emph{right}). \emph{(b)} Persistence diagram of $DSI^{cort}$ and $MW^{cort}$. \new{Topological cavity} in \emph{(a)} circled in maroon. \emph{(c)} Patterns of connectivity between maroon \new{loop} nodes found for the original \emph{(c)} and contralateral \emph{(d)} hemispheres in each scan. If the exact pattern is not found, the pattern at the edge density when all cycle edges first exist is shown. For each scan, the connection pattern of nodes in the minimal generator with the fewest number of cross-edges is shown.}
	\label{fig:sfig_1cyclesub}
	
\end{figure}

Densely connected subcortical nodes \new{may prevent the longevity of nonzero homology classes} by forming cross-cycle edges or \new{cliques which tessellate the} cycle completely. We asked what \new{cavities} could be found when removing these subcortical nodes, forming $DSI^{cort}$ as described in the main text. Here, Fig.~\ref{fig:sfig_1cyclesub} shows a 1-cycle on nine nodes recovered from $DSI^{cort}$ within the brain and as a schematic (panel \emph{(a)}). The persistence diagram for \new{2D cavities} within $DSI^{cort}$ in Fig.~\ref{fig:sfig_1cyclesub}b shows \new{the four minimal cycles} marked in maroon. Importantly, because of the connection patterns between nodes at the density of cycle formation, \new{we will refer to any of these four cycles as the minimal cycle}. Two of these \new{cycles} are equivalent loops which involve the superior frontal (RH) and the caudal middle frontal regions. The other two are equivalent to each other but not to the first two loops, and involve the superior frontal (LH) and posterior cingulate (LH). The edge added at $\rho_{birth}$ connects the lateral orbitofrontal to the superior temporal. \new{The cycle formed by the superior frontal (RH, LH), caudal middle frontal, precentral, and posterior cingulate (LH) is itself a minimal cycle surrounding a separate topological cavity.} This information along with the connection patterns at $\rho_{birth}$ mean we cannot claim either pair are the two minimal generators, instead it is either one pair or the other. The smaller, five node cycle was already in existence, so either of these possible paths (but not both simultaneously) completes the larger maroon cycle.

We see the pattern of connectivity is not often exactly seen in all individuals, yet the large 2-dimensional cavity enclosed is present in every scan (Fig.~\ref{fig:sfig_1cyclesub}c) in the original hemisphere, and often in the opposite hemisphere (Fig.~\ref{fig:sfig_1cyclesub}d), suggesting its importance in neural structure.  

\begin{figure}[h!]
	\centering
	\includegraphics[width = 3.5in]{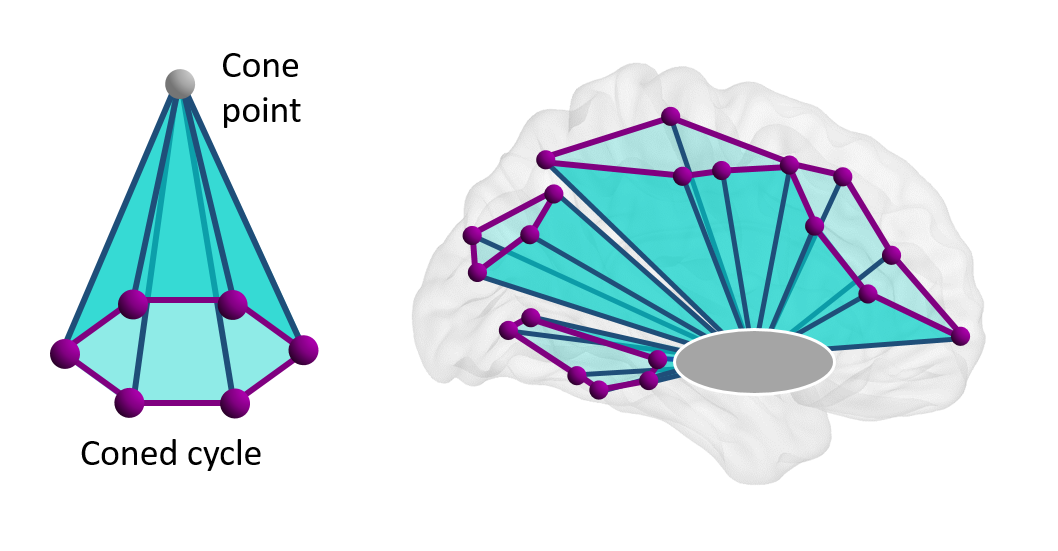}
	\caption{\textbf{Subcortical regions as cone points in the brain network.} A loop (maroon, \emph{left}) may be the base of a cone, where the cone point (gray) triangulates the loop interior thus making the loop a boundary loop. In the brain, the \new{greater number and longevity of topological cavities} seen after removing subcortical nodes indicates these subcortical regions (gray, \emph{right}) may act as cone points for many cortical cycles.}
	\label{fig:sfig_cone}
\end{figure}

The number and pattern of persistent cycles in Fig.~\ref{fig:sfig_1cyclesub}b matches that of the minimally wired null model much more closely than the full DSI network. Thus suggests first that the cortical wiring of the brain is globally arranged as if it was wired minimally. Yet the difference in the cortical only and full DSI persistence diagrams also implies the subcortical regions drive the reduction of homology. Knowing the subcortical regions are highly connected and participate in many high-dimensional cliques (Fig.~\ref{fig:fig2}), we conclude the subcortical regions are acting as cone points in the brain network (Fig.~\ref{fig:sfig_cone}, left). Finally, this adds more detail to our understanding of the global wiring of the brain, as we imagine many cortical loops that are coned by sets of subcortical regions (Fig.~\ref{fig:sfig_cone}, right).

\subsection{Persistent homology of an alternative parcellation}

\begin{figure}[h]
	\centering
	\includegraphics[width = 5in]{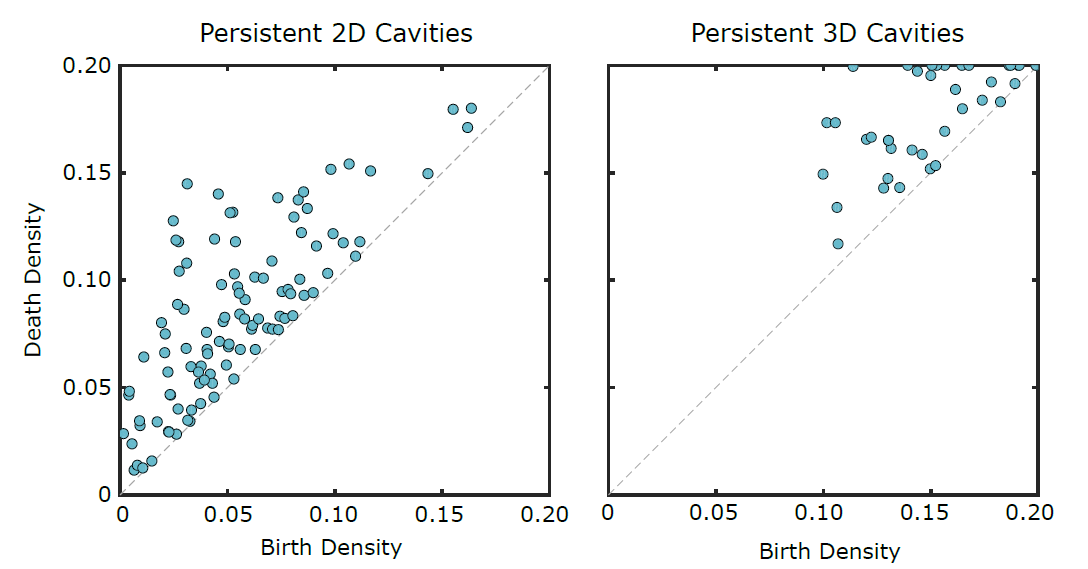}
	\caption{\new{Persistence diagram of the structural connectome with the AAL 90 parcellation.}}
	\label{fig:aal90}
\end{figure}

\new{Throughout this investigation we have exclusively used the 83-node Lausanne parcellation \cite{cammoun2012mapping} for consistency. Yet, it is important to consider how the choice of parcellation might impact on the observed architectures. Currently, there is no single agreed upon parcellation scheme for either structural or functional imaging data. Moreover, it is not yet agreed upon whether separate parcellations should be used for the two sorts of imaging data. In this work, we have used a parcellation scheme that is derived from the subject-level rescontruction of the gray matter surface, based on the state-of-the-art application of FreeSurfer software. Thus, our parcellation is one that should maximize sensitivity to individual differences in surface morphology, and thus maintain optimal identification of boundaries between morphological features. Other parcellation techniques are available -- such as the registration of an averaged template to the subject volume. However, these techniques have the potential disadvantage of enforcing a group-level solution for parcels onto a subject-specific image. Nevertheless, for completeness, we also examined such an alternative parcellation scheme -- by applying th 90-region AAL atlas to the subject-level images. Shown in Fig.~\ref{fig:aal90}, we observed more persistent features than in the original 83-node parcellation, both in dimensions one and two, consistent with the fact that in networks with geometric constraints, we expect the number of features to increase with more nodes \cite{giusti2015clique}. Additionally, we note this persistent homology signature is more geometric \cite{kahle2011random} than the original network (Fig.~\ref{fig:fig4}). Importantly, exact comparisons between the cycles in one parcellation and those in the other parcellation is intractable because the two parcellations do not contain the same number of regions, or same anatomical location of regions.}

%
%
%

\end{document}